\documentclass[authoryear,12pt]{article}
\usepackage{amsmath}
\usepackage{graphicx}
\usepackage{amsfonts}%
\usepackage{amssymb}
\usepackage{floatrow}
\usepackage{times}
\usepackage{array}
\usepackage{blindtext}
\usepackage[]{apacite}
\RequirePackage[colorlinks,citecolor=blue,urlcolor=blue]{hyperref}

\usepackage{natbib}

\usepackage{subcaption}

\allowdisplaybreaks

\usepackage{enumitem}
\usepackage{amsthm}
\usepackage{setspace}
\usepackage{caption}
\usepackage[margin=1in]{geometry}

\newcommand{\bet}{ \mbox{\boldmath $ \eta $} }

\newcommand{\bbet}{ \mbox{\boldmath $ \beta $} }
\newcommand{\bbeta}{ \mbox{\boldmath $ \beta $} }

\newcommand{\bzeta}{ \mbox{\boldmath $\zeta$}}

\newcommand{\beps}{ \mbox{\boldmath $\epsilon$}}
\newcommand{\bepsilon}{ \mbox{\boldmath $\epsilon$}}

\newcommand{\bnu}{ \mbox{\boldmath $\nu$} }
\newcommand{\bmu}{ \mbox{\boldmath $\mu$} }

\newcommand{\bSigma}{ \mbox{\boldmath $\Sigma$} }

\newcommand{\bLam}{ \mbox{\boldmath $\Lambda$} }
\newcommand{\bLambda}{ \mbox{\boldmath $\Lambda$} }

\newcommand{\bgamma}{ \mbox{\boldmath $\gamma$} }

\newcommand{\boeta}{ \mbox{\boldmath $\eta$} }
\newcommand{\bomega}{ \mbox{\boldmath $\omega$} }

\newcommand{\bzero}{\textbf{0}}

\newcommand{\bZ}{\textbf{Z}}
\newcommand{\bz}{\textbf{z}}

\newcommand{\bA}{\textbf{A}}

\newcommand{\bB}{\textbf{B}}

\newcommand{\bD}{\textbf{D}}

\newcommand{\bI}{\textbf{I}}

\newcommand{\bG}{\textbf{G}}

\newcommand{\bm}{\textbf{m}}

\newcommand{\bV}{\textbf{V}}

\newcommand{\bx}{\textbf{x}}
\newcommand{\bX}{\textbf{X}}

\newcommand{\bY}{\textbf{Y}}

\newcommand{\bR}{\textbf{R}}

\newcommand{\pmt}{$\text{PM}_{10}$ }

\defcitealias{data}{SEDEMA, 2017}

\title{Multivariate Functional Data Modeling with Time-varying Clustering}
\author{Philip A. White and Alan E. Gelfand}

\begin{document}

\maketitle

\begin{abstract}

We consider the situation where multivariate functional data has been collected over time at each of a set of sites.  Our illustrative setting is bivariate, monitoring ozone and PM$_{10}$ levels as a function of time over the course of a year at a set of monitoring sites.  Our objective is to implement model-based clustering of the functions across the sites.  Using our example, such clustering can be considered for ozone and PM$_{10}$ individually or jointly.  It may occur differentially for the two pollutants.  More importantly for us, we allow that such clustering can vary with time.

We model the multivariate functions across sites using a multivariate Gaussian process. With many sites and several functions at each site, we use dimension reduction to provide a stochastic process specification for the distribution of the collection of multivariate functions over the say $n$ sites. Furthermore, to cluster the functions, either individually by component or jointly with all components, we use the Dirichlet process which enables shared labeling of the functions across the sites. Specifically, we cluster functions based on their response to exogenous variables.  Though the functions arise in continuous time, clustering in continuous time is extremely computationally demanding and not of practical interest. Therefore, we employ a partitioning of the time scale to capture time-varying clustering.

The data we work with is from 24 monitoring sites in Mexico City which record \emph{hourly} ozone and PM$_{10}$ levels.  We use the data for the year 2017.  Hence, we have 48 functions to work with.  We provide a Gaussian process model for each function using continuous time meteorological variables as regressors along with adjustment for daily periodicity.  We interpret similarity of functions in terms of their shape, captured through site-specific coefficients and use these coefficients to develop the clustering.

%

\end{abstract}

\section{Introduction}\label{sec:intro}

The beginnings of functional data analysis date to
\cite{grenander1950} and \cite{rao1958}, who
applied the theory of stochastic processes to curve data. The term ``functional data analysis'' came into common usage with \cite{ramsay1982} and \cite{ramsay1991}. Since then the field has undergone rapid
growth, spurred by the re-framing of existing problems
in the context of functional data analysis along with scientific advances giving rise to larger amounts and new types of functional
data. Early work focused mainly on curves as the object of study. \cite{ramsay2005} and \cite{ramsay2007} provide a thorough introduction to the basic concepts
of functional data analysis. Applications have been found in MRI brain images (and, in fact, for imaging in general), finance, climatic variation, spectrometry data, and time-course gene expression
data, to name a few. For a comprehensive overview of applications, see \cite{ullah2013}.

Functional data is viewed explicitly as
living on $\mathbb{R}^d$, where $d=1,2,3$ in most applications.  The response is indexed by arguments that are viewed as
\emph{continuous}.  Of course, for any function, samples are only ever collected at a finite set of arguments.
In fact, finite-dimensional multivariate approaches were used
originally for FDA.

Explicit modeling of functional data is usually carried out in one of
two ways: (i) as linear combinations of some set of basis functions, or
(ii) as realizations of some stochastic process. In the former case, weights are chosen for the basis functions so that the fit to the data is
optimized (\cite{ramsay2007, morris2015} ). For instance, splines
implement this by imposing a fixed partition on the domain of the
function and allowing for the weights to vary across the partition
elements. Observed data are then viewed as noisy realizations of the
underlying splines, with implementation requiring finite truncation of
the basis function expansion.  To address the high
dimensionality of functional data, we find extensions of principal component
analysis to the functional domain.  Formally, functional principal components
analysis (FPCA) specifies each function as an infinite linear
combination of orthonormal basis functions using the
Karhunen-Lo\'{e}ve
Theorem. Practically, a finite linear combination is adopted.  Arguably, FPCA is among the most popular techniques for analyzing
functional data.

The alternative approach which we pursue here is to view the data for each individual as
noisy observations from a realization of a stochastic process.  That is, the \emph{true}
functions are realizations of a stochastic process over the subset of
$\mathbb{R}^{d}$ of interest.  Gaussian processes are often a convenient choice since
they require only the specification of a mean function and a valid
covariance function.  They are flexible and provide convenient
standard multivariate normal distribution theory as well as straightforward interpolation.  Extensions of this approach, based upon Dirichlet process modeling ideas, developed below,  enable our functional modeling to exhibit clustering which is the primary contribution of this paper.

With multivariate functional data, methods employed for clustering of
multivariate data vectors can be adapted to the task of clustering functional data, with the usual goal
being to group observations such that within-cluster
distance under some criterion is minimized. For example, K-means clustering is
a widely applied algorithm for this problem \citep{hartigan1979, abraham2003}.  Here, we avoid an algorithmic approach and, instead, pursue model-based clustering which can be placed under the heading of mixture modeling. At a ``big picture''  level, the different mixture
components act as the cluster centers, and optimization is equivalent
to the model fitting process \citep{sugar2003, jacques2014}. We note that this approach addresses clustering for univariate functional data; see \cite{wang2016} for a comprehensive review.

Multivariate  functional  data  typically comprise  several  simultaneously
recorded time course measurements for a sample of subjects or experimental
units. They are viewed as realizations sampled from multivariate random functions. Analysis of multivariate functional data has
received relatively less attention than univariate functional data methods though chemometrics is an area where multivariate FDA has been discussed \citep[see, e.g.,][]{wang2015}..
Again, functional principal component analysis (FPCA) serves as a fundamental tool.
Methods for analyzing multivariate functional data that utilize univariate
FPCA include dynamical correlation
analysis for multivariate longitudinal observations \citep{dubin2005}, modeling the relationship of paired longitudinal observations \citep{zhou2008}, and regularized FPCA for multidimensional functional data
\citep{kayano2009}.
When the components of multivariate functional data are measured in the
same units and have similar variation, classical multivariate FPCA that concatenates the multiple functions into one to perform univariate FPCA can work well.

The approach we adopt here employs multivariate Gaussian process specification for the functions. \cite{shi2011} presents discussion in this vein.  In particular, we consider the situation where multivariate functional data has been collected over time (so $d=1$) at each of a set of sites.  Our illustrative setting is bivariate, monitoring ozone and PM$_{10}$ levels as a function of time over the course of a year, at a set of monitoring sites.  Our objective is to implement model-based clustering of the functions across the sites.  Using our example, such clustering can be considered for ozone and PM$_{10}$ individually or jointly.  It may occur differentially for the two pollutants.  More importantly for us, we allow that such clustering can vary with time.

Specifically, we consider the challenge of modeling collections of multivariate functional data incorporating their relationships to exogenous variables. We consider the setting where we have multiple functional outcomes that are related by regression to predictors that are themselves functional, up to pure error.  In addition, mean $0$ random effects are introduced using Gaussian processes, one for each function at each site.  These random effects provide local (in time) adjustment to the regression functions.

In this setting, we interpret clustering to signify a common functional form for the functions, ignoring the random effects. If our functions are specified as Gaussian processes with a mean that is a regression in exogenous variables, then we cluster two functions if they share the same regression coefficients.  In this regard, we can distinguish between the ``shape'' of the function and the ``level'' of the function.  In our simple setting, sharing shape is captured through common slope coefficients; sharing level is captured by additionally sharing the intercept.  Most importantly, we \emph{don't} cluster functions in terms of response.  That is, the various functions may be observed at quite different levels of the exogenous variables resulting in quite different observed response levels.  Again, we seek to cluster functions which share shape.

We are fortunate that all of our functions run for the same length and period of time.
So, we can avoid some of the well-discussed challenges in the literature regarding registration
and time warping in order to ``align'' the functions (see \cite{gervini2014} for a general discussion and see \cite{telesca2008} within the Bayesian framework).


In our context, we have $n=24$ sites with two functions at each site.  Each function is observed hourly over a year, resulting in $8760$ observations for each.  We seek to capture dependence between the functions within a site as well as dependence between the functions across the sites.  We use dimension reduction to specify the joint distribution of the collection of multivariate functions over the $24$ sites, expressing the $2 \times 24$ functions in terms of $2 \times r$ functions, where, illustratively, $r$ might be $3$ or $5$.  We model these $2r$ functions using multivariate Gaussian processes which induce a Gaussian process specification for each function at each site.

With regard to our clustering objective, it is clear that two Gaussian process realizations over a subset say $(0,T]$ of $R^1$ will agree with probability $0$.  So, in order to achieve model-based clustering, we imagine a finite or countably infinite set of functional realizations over $(0,T]$.  Then, two functions are clustered if they share a common realization from this set. That is, we need to add a further level of hierarchical modeling to provide the set of functional realizations which are available to share.  A convenient way to do this is through the Dirichlet process (which is discussed and referenced below).  Such processes provide random discrete distributions, i.e., a random set of atoms with random probabilities assigned to the atoms.  In our setting the random atoms are functions.  When functional data is modeled in this fashion, two functions are clustered if they share the same atom, i.e., each atom is assigned a label and the two functions receive the same label.

In fact, there is more subtlety here according to whether or not the functions are explained through the introduction of regressors.  If no regressors are considered, then all of the atoms are realizations of a constant mean Gaussian process over $(0,T]$ whence all of the functions we are modeling are constant mean Gaussian processes.  If regressors are introduced into the means of the Gaussian processes, then we can consider two functions as clustered if they share the same mean function, ignoring the mean $0$ residual (local adjustment) process.  Specifically, with a regression function of the form say $\bX(t) \bbeta_{i}^{(k)}$ for the $k$th function at site $i$, clustering involves sharing of coefficient vectors for the $\bbeta_{i}^{(k)}$.  Now, the Dirichlet process would introduce atoms which are vectors of regression coefficients.  We adopt this latter approach here since we seek to understand the nature of differential pollutant response to meteorological predictors across the sites.

We add a further wrinkle.  We imagine that the relationship between response and predictors is changing over time.  In particular, we use a partitioning of the time scale to capture time-varying change in these response/predictor relationships.  In turn, this leads to an investigation of time-varying clustering of the functions.  For instance, what can we say about how clustering of the sites arises say in summer vs. in winter?  Though the functions arise in continuous time, continuous time clustering is not of practical interest and, additionally, is extremely computationally demanding.  Instead, illustratively, we work at monthly scale.

Finally, though the sites are spatially referenced, we intentionally do not model them spatially.  We are not interested in interpolation of the functions.  Though there may be a conceptual realization of a multivariate function at every spatial location, we don't imagine clustering functions which we don't observe.  In fact, we seek to cluster functions through their functional similarity which is not a spatial notion.  As a post model fitting exercise, we can assess how strong clustering is in terms of inter-site distance.



The data we work with is from 24 monitoring sites in Mexico City which record hourly ozone and PM$_{10}$ levels.  We use the data for the year 2017.  We have 48 functions to work with and we provide a Gaussian process model for each function using continuous time meteorological variables - relative humidity and temperature -  as regressors along with daily periodicity to supply the mean of each of the Gaussian processes.  The model fitting provides site-specific coefficients and we use these coefficients to develop the clustering.
Again, since the data at each site is collected over the same window of time, no alignment techniques are needed.  Furthermore, the fact that these stations measure the same quantities and are close in proximity suggests that the data can be effectively modeled using a representation of lower rank than the 24 pairs of curves we observe.

The format of the paper is as follows. We continue with a review of the modeling components we utilize in Section \ref{sec:review}, focusing on multivariate Gaussian processes (Section \ref{sec:mgpf_intro}) and Dirichlet processes (Section \ref{sec:dp_intro}). We present our modeling framework in Section \ref{sec:methods_models}, including a discussion of prior distributions and model fitting. In Section \ref{sec:mexico_city}, we present our analysis of ozone and \pmt in Mexico City in year 2017. We discuss data attributes in Section \ref{sec:data}, specific modeling details in \ref{sec:mexico_city_model}, and results of our analysis in Section \ref{sec:results}. Lastly, in Section \ref{sec:conclusions}, we summarize our approach and discuss possible extensions.

\section{The Model Components}\label{sec:review}

The general form of the modeling is as follows.  At site $i$ for function $k$ at time $t$, we assume
\begin{equation}
Y_{i}^{(k)}(t) = f_{i}^{(k)}(t) + \epsilon_{i}^{(k)}(t)
\end{equation}
where $Y$ denotes the observed response, $f$ denotes the ``true'' function and $\epsilon$ denotes pure error.  Further, we express the function in terms of a ``fixed'' effects term, $\mu_{i}^{(k)}(t)$ and a ``random'' effects term, $\eta_{i}^{(k)}(t)$, i.e.,
\begin{equation}
f_{i}^{(k)}(t) = \mu_{i}^{(k)}(t) + \eta_{i}^{(k)}(t).
\end{equation}
Here, $\mu_{i}^{(k)}(t)$ is a regression term which, illustratively, we specify as $\bX_{i}(t)^{T}\bbeta_{i}^{(k)} + \bZ_{i}(t)^{T}\bgamma_{i}^{(k)}$, i.e., a component and site specific vector of regression coefficients. Here, we let $\bbeta_{i}^{(k)}$ be the regression coefficients that are involved in clustering and $\bgamma_{i}^{(k)}$ are the regression coefficients not involved in clustering allowing for settings where clustering may be of interest for only a subset of covariates.  $\eta_{i}^{(k)}(t)$ is a mean 0 Gaussian process and these processes are \emph{dependent} across $i$ and across $k$.  We can assemble the model at site-level to
\begin{equation}
\bY_{i}(t) = \mathbf{B}_{i} \bX_{i}(t) + \mathbf{G}_i \bZ_i(t) + \boldsymbol{\eta}_{i}(t) + \beps_{i}(t).
\label{eq:gen}
\end{equation}
Here, with $n$ sites, $i=1,2,...,n$.  With $K$ functions at each site and $p$ predictors (including intercept), $\mathbf{B}_{i}$ and $\mathbf{G}_{i}$ are $K \times p$ with $k$th row, $(\bbeta_{i}^{(k)})^{T}$ and $(\bgamma_{i}^{(k)})^{T}$, respectively.  $\boldsymbol{\eta}_{i}(t)$ is a $K$-dimensional Gaussian process and the $\boldsymbol{\eta}_{i}(t)$ are dependent across $i$.  So, in total, we have $nK$ functions specified as an $nK$ dimensional Gaussian process.  Finally, the $\beps_{i}(t)$ are pure error vectors, independent across components, and normally distributed with mean $0$ and $k$th component having variance $\tau^{2(k)}$.

Equation (\ref{eq:gen}) looks like a familiar mixed effects model for continuous time data.  However, we introduce the following novelties.  First, we partition the time window into say $M$ segments, i.e., $(0, t_1, t_2,..., t_{M-1},T]$.  On the $m$th segment we revise (\ref{eq:gen}) to
\begin{equation}
\bY_{i}^{m}(t) = \mathbf{B}_{i}^{m} \bX_{i}(t) + \mathbf{G}_{i}^{m} \bZ_{i}(t) + \boldsymbol{\eta}_{i}(t) + \beps_{i}(t), \hspace{.2cm} t_{m-1} < t \leq t_{m}.
\label{eq:timevar}
\end{equation}
As a result, we have time-varying coefficients.  However, as residuals, we assume that the random effects Gaussian processes and the pure error processes are not partition-dependent.

Second, we implement a dimension reduction strategy to model the $\boldsymbol{\eta}_{i}(t)$ using an $rK$ dimensional Gaussian process where $r$ is much smaller than $n$, typically, say $3$ to $5$.  Since the number of temporal observations for each function is large (hourly over an entire year in our case), even with independent Gaussian process, we will wind up working with very high dimensional covariance matrices for each process. For likelihood evaluation, these matrices require inverse and determinant calculation.  There are strategies for handling such dimensionality challenges such as the predictive process \citep{banerjee2008} or the nearest neighbor Gaussian process \citep{datta2016a}.  However, since our processes are over time, a convenient way to avoid this computational challenge is to adopt an exponential covariance function.  This corresponds to a continuous time AR(1) process \citep[see, e.g.,][]{brockwell2007}.  With equally spaced time points, this enables writing of the joint distribution sequentially over time without resorting to calculating the joint covariance matrix.

Third, we recall our objective of clustering.  With time-varying coefficients for each component function, we can investigate clustering of the components across the sites within each partition.  We can investigate whether and, if so, how clustering behavior changes with partition, i.e., with time.  As discussed in the Introduction, the Dirichlet process provides a convenient distributional mechanism to enable ties, i.e., to enable, within partition $m$, $\bbeta_{i}^{m(k)} = \bbeta_{i'}^{m(k)}$ for one or more $k$'s.

\subsection{Multivariate Gaussian Process Factor Models}\label{sec:mgpf_intro}

The multivariate Gaussian process (GP) supplies both a data model for multivariate data collected at continuously indexed \emph{sites} as well as a multivariate random effects process model when dependence between the processes is sought.  For examples of the first setting see, e.g., \cite{wackernagel1994,gneiting2010,vandenberg2015}.  For the second, see, e.g., \cite{banerjee2014}, Chap 9.
Regardless, specification of a multivariate Gaussian process requires a valid cross-covariance function. In the above notation, indexing over time, we seek a site specific $K \times K$ matrix valued function, $C_{i}(t, t')$ with $(C_{i}(t,t')_{k,k'} = \text{cov}(\eta_{ik}(t), \eta_{ik'}(t'))$. There are numerous constructions for $C$ including coregionalization, kernel convolution, and convolving covariance functions \citep[][Chap 9]{banerjee2014}. Direct development has been pursued for the Mat\'ern covariance class in, e.g., \cite{gneiting2010,apanasovich2012}.  Development through stochastic partial differential equations is presented in, e.g., \cite{vandenberg2015,singh2018}.  For convenience, particularly since, in our application $K=2$, we employ coregionalization.  However, any of the foregoing choices could be used equally well.

As above, we seek to have the $\bet_{i}(t)$ dependent across $i$.  We want to specify an $nK$ dimensional GP through a dimension reduction which requires only an $rK$ dimensional GP.  That is, for each $i$, we want to specify $\eta_{i}^{(k)}(t) = \Lambda_{i}^{(k)T} \bomega^{(k)}(t)$ where $\Lambda_{i}^{(k)}$ is an $r \times 1$ vector and $\bomega^{(k)}(t)$ is an $r$-dimensional GP with mean $\mathbf{0}$ and correlation matrix $\bR$.  Assembling across $n$, we obtain the $n \times 1$ vector, $\bet^{(k)}(t)$ such that
\begin{equation}\label{eq:func_factor}
\bet^{(k)}(t)= \left(
                 \begin{array}{c}
                   \eta_{1}^{(k)}(t) \\
                   \eta_{2}^{(k)}(t) \\
                   . \\
                   . \\
                   . \\
                   \eta_{n}^{(k)}(t) \\
                 \end{array}
               \right) = \bLam^{(k)}\bomega^{(k)}(t)
\end{equation}
where $\bLam^{(k)}$ is an $n \times r$ matrix.  What we have here is a latent factor analysis \citep{lopes2004}  where $\bLam^{(k)}$ provides the matrix of factor loadings for the $k$th component of the $\bet_{i}(t)$ vectors.  We use correlation functions (covariance functions with unit variance) in the $\bomega^{(k)}(t)$ above in order to identify the factor loadings. We emphasize that the dimension reduction is across sites, not across components of $\bet_{i}(t)$.  Specifically, we provide this dimension reduction without introducing the spatial locations of the sites.  In fact, $\bLam^{(k)}\bLam^{(k)T}$ provides the $n \times n$ covariance matrix between the sites.  In order to see how dependence for the $k$th component associates with distance between sites, we will investigate the $(\bLam^{(k)}\bLam^{(k)T})_{i,i'}$ vs. the corresponding inter-site distances, $d_{i,i'}$.

To complete the specification we need to provide the modeling for the set of $rK$ processes, $\omega_{j}^{(k)}(t)$, $j=1,2,...,r; k=1,2,...,K$.  Here, we bring in coregionalization, writing the $K \times 1$ vector, $\bomega_{l}(t)$ as
\begin{equation}
\bomega_{l}(t) = \left(
                   \begin{array}{c}
                     \omega_{l}^{(1)}(t) \\
                     \omega_{l}^{(2)}(t) \\
                     . \\
                     . \\
                     . \\
                     \omega_{l}^{(K)}(t) \\
                   \end{array}
                 \right) = \bA \bnu_{l}(t) =  \bA \left(
                   \begin{array}{c}
                     \nu_{l}^{(1)}(t) \\
                     \nu_{l}^{(2)}(t) \\
                     . \\
                     . \\
                     . \\
                     \nu_{l}^{(K)}(t) \\
                   \end{array}
                 \right)
\end{equation}
where $\bA$ is a $K \times K$ coregionalization matrix and the $\nu_{l}^{(k)}(t)$ are i.i.d mean $0$, variance $1$ GPs.  In particular, by adopting an exponential correlation function, we assume they are continuous time first order Markov processes.  Hence, the joint distribution across time can be peeled off sequentially, i.e.,
\begin{equation}
[\nu_{l}^{(k)}(t_1), \nu_{l}^{(k)}(t_2),..., \nu_{l}^{(k)}(t_{s-1}), \nu_{l}^{(k)}(t_{s}),...]
\end{equation}
becomes
\begin{equation}\label{eq:ar_ou}
[\nu_{l}^{(k)}(t_1)][\nu_{l}^{(k)}(t_2)|\nu_{l}^{(k)}(t_1)]...[\nu_{l}^{(k)}(t_{s})|\nu_{l}^{(k)}(t_{s-1})]...
\end{equation}
where $[\nu_{l}^{(k)}(t_{s})|\nu_{l}^{(k)}(t_{s-1})] = N(e^{-\phi_{(k)}(t_{s} - t_{s-1})} \nu_{l}^{(k)}(t_{s-1}),1 -e^{-2 \phi_{(k)} (t_s - t_{s-1}) })$ with $[\nu_{l}^{(k)}(0)] = N(0,1)$.

Again, the factor loadings for the $k^\text{th}$ component are captured in the $n \times r$ matrix $$ \bLambda^{(k)} = \left( {\bLam^{(k)}_1},...,{\bLam^{(k)}_{n}} \right)^T ,$$ which brings in the dependence across the sites. The coregionalization matrix $\bA$ induces dependence between the random effects at a given site.  Since coregionalization is applied to random effects, for convenience we can take $\bA$ to be lower triangular.

We recall that the factor model is not identifiable and thus constraints to $\bLam$ are applied to remedy this. It is common to regain identifiability of $\bLam$ by either (i) constraining $\bLam$ to be orthogonal \citep[e.g.][]{seber2009} or lower triangular \citep[see][]{geweke1996,aguilar2000,lopes2004} or (ii) using prior distributions that induce sparseness or that provide shrinkage \citep[see][for example]{bernardo2003,bhattacharya2011}. Identifiability can also be recovered by constraining parameters of the GPs. For example, the range parameters of the GPs could be ordered to make the factor loading identifiable. In this case, the factors would be interpretable in the way they each account for different ranges of autocorrelation. We discuss this approach in our model fitting discussion in Section \ref{sec:priors_fitting} and in our analysis of hourly pollution levels presented in Section \ref{sec:mexico_city}.



We conclude this subsection by noting that our specification enables convenient separation of the factor loadings and the coregionalization entries in the dependence structure for the $\eta_{i}^{k}(t)$.  Specifically, for say $t' \geq t$, we have

\begin{equation}
\text{cov}(\eta_{i}^{k}(t), \eta_{i'}^{k'}(t') = (\bLam_{i}^{(k)})^{T}\bLam_{i'}^{(k')}\sum_{j=1}^{K}A_{kj}A_{k'j}e^{-\phi_{(j)}(t'-t)}.
\label{eq:dep}
\end{equation}
We see the separation of the factor loadings at $i$ and $i'$ for $k$ and $k'$ from the coregionalization at $k$ and $k'$.  The proof of result \eqref{eq:dep} is provided in the Appendix.
Following from this, we can directly obtain all of the marginal and conditional distribution theory needed for the model fitting.  These results are also supplied in the Appendix.


\subsection{Dirichlet Processes}\label{sec:dp_intro}

Bayesian hierarchical specifications provide a convenient way to develop mixture
models. Specifically, the Dirichlet process has been used to avoid explicitly
specifying the number of clusters present in the data. That is, we need not struggle with the identifiability and slow-mixing challenges of finite mixture models.  The Dirichlet process provides a countable mixture model where, again, the intent is clustering rather than learning about the number of mixture components and the parameters associated with each component.  In this regard, the Dirichlet process model serves as a prior assigning random distributions which have countably infinite support; it dates at least to (Ferguson, 1973).  The stick-breaking
representation presented by (Sethuraman, 1994) consequentially expanded its applicability.
Model fitting using Markov chain Monte Carlo (MCMC) was initially
presented in (Escobar and West, 1995). In particular, Dirichlet process mixtures of
stochastic processes are possible, allowing for extensions to accommodate functional data.

The random discrete distributions can place mass on any random objects of interest.  Hence, they can adopt as functional atoms, realizations of Gaussian processes or realizations of spline-specified functions \citep[see, e.g.,][]{maceachern2000,gelfand2005,duan2007,petrone2009}.  However, as we noted in the Introduction, here, clustering corresponds to sharing regression coefficient vectors since, across sites, we seek to cluster pollutant response to meteorological predictors.

Following \cite{ferguson1973}, Dirichlet processes are defined in terms of a base distribution
$G_{0}$ along with a concentration parameter $\alpha$; we say that a realization $G $ from a Dirichlet process is a random, almost surely discrete distribution
that is close, in some sense, to the base measure $G_{0}$ with the closeness determined by
the concentration parameter $\alpha$ and we write $G\sim DP(\alpha, G_{0})$.

The constructive representation introduced by Sethuraman (1994), showed
that countably infinite mixtures of atoms with suitable stick-breaking weights are
equivalent to draws from a Dirichlet process. The base measure of the Dirichlet
process is then the distribution that provides the mixture components in the infinite
mixture. That is, $G \sim DP(\alpha, G_{0})  \Longleftrightarrow  G = \sum_{l=1}^{\infty} p_{l} \delta_{\theta_{l}^{*}}$ where the $\theta_{l}^{*}$ are i.i.d. realizations from $G_{0}$ and the $p_{l}$
arise as stick-breaking probabilities, i.e., $p_1 = w_1$ and $p_{l} = w_{l}\Pi_{j=1}^{l-1}(1-w_{j})$ with the $w_{l} \sim \text{Beta}(1, \alpha)$.

The stick-breaking representation provides intuition for sampling
schemes. The mixture representation suggests introducing ``latent classes''
that individual observations are assigned to, resulting in distributions that are easier
to sample. This enables simplified Markov chain Monte Carlo (MCMC) model fitting.  In practice it is customary to truncate the number of $p_{l}$'s such that $\sum_{l=1}^{L} p_{l}$ is suitably close to $1$ or, equivalently, $\Pi_{l=1}^{L} (1-w_{l})$ is sufficiently close to $0$. Alternatively, the Dirichlet process can be represented be an urn scheme \citep{blackwell1973}. The Gibbs sampler for this approach is straightforward and presented in \cite{escobar1994,escobar1995}.
Using the P\'{o}lya-urn representation of the Dirichlet process, the model for data $\bY = \left(y_1,...,y_n \right)$ is represented
\begin{align}
\label{eq:model_DP_polya}
p(\bY) &= \prod^n_{i = 1} g(y_i | \zeta_i) \\
\pi(\bzeta) &= \prod^n_{i = 1} \left[ \frac{\alpha}{\alpha + i  - 1}  G_0(\zeta_i) + \frac{1}{\alpha + i - 1} \sum_{l < i} \delta_{\zeta_l}\left(\zeta_i \right) \right] \nonumber.
\end{align}
This representation reduces the infinite mixture to (at most) an $n$-dimensional mixture, where $\bzeta = (\zeta_1,...,\zeta_n)^T$ assigns membership to specific mixture components. The full conditional for each $\zeta_i | \zeta_{-i}$ is
\begin{equation}\label{eq:full_conditional_DP_polya}
\pi(\zeta_i | \zeta_{-i}) =  \frac{\alpha}{\alpha + n  - 1}  G_0(\zeta_i) + \frac{1}{a + n - 1} \sum_{l < i} \delta_{\zeta_l}\left(\zeta_i \right) .
\end{equation}
The representations in (\ref{eq:model_DP_polya}) and (\ref{eq:full_conditional_DP_polya}) indicate that with more $\zeta_i$ assigned to the same mixture component, it becomes more likely that other observations will be assigned to that component.

In our setting, we use a DP mixture prior distribution on function-specific regression coefficients to identify unique responses to exogenous variables or predictors. Heuristically, this prior distribution groups functional data that have similar responses to predictors by inducing clusters or ties between regression coefficients.  The DP-mixing accrues to the functions $Y_{i}^{(k)}(t)$'s.  That is, each component at each site has a random coefficient vector, $\bbet_{i}^{(k)}$.

Using the Dirichlet process, these $\bbet_{i}^{(k)}$ can be modeled jointly across $k$ or individually for each $k$ according to whether we seek to cluster by site or by components within site.  In either case, under the Dirichlet process, each $\bbet_{i}^{(k)}$ receives a random label.  For joint clustering, we denote it as $L_{i}$ and if $L_{i}=l$ then we set $B_{i} = B_{l*}$, the $l$th atom in the distribution of $G$.  If it is by component, then we denote it as $L_{i}^{(k)}$ and if $L_{i}^{(k)} = l$,  $\bbet_{i}^{(k)} = \bbet_{l}^{(k)*}$, now, the $l$th atom in the distribution say, $G^{(k)}$ for the $k$th component.  The DP-mixing occurs because $Y_{i}^{(k)}(t)|\bbet_{i}^{(k)}$ is normally distributed, i.e., the $Y_{i}^{(k)}(t)$ arise from a mixture distribution given the $\bbet_{l}^{(k)*}$ atoms.  Articulating the specification in this fashion also reveals that we primarily care about when two sites share labels with little interest in the values of the associated atoms.  Atoms will change across iterations of MCMC model fitting and so will labels.

Lastly, we allow this labeling to vary over time using exchangeable DP mixtures as prior distributions over coarse time scales. This requires introducing a superscript $m$ in each of the quantities in the previous paragraph.  In some settings, incorporating dependence between the base measures for the time-varying DP mixtures may be warranted (e.g., the mean of the base measure may depend on the mean over the previous time period).  Fitting such models is very challenging and is beyond the scope of our work here.

\section{Methods and Models}\label{sec:methods_models}


We now return to the model in (4) which allows for partitioning of the time scale.  Our data is still denoted by $Y_{i}^{(k)}(t)$.  This model, at the component scale, becomes
\begin{equation}
Y_{i}^{(k)}(t) = \bX_{i}(t)^T \bbeta_{i,m}^{(k)} + \bZ_{i}(t)^T \bgamma_{i,m}^{(k)} + \eta_{i}^{(k)}(t) + \epsilon_{i}^{(k)}(t), \hspace{.2cm} t_{m-1} < t \leq t_{m}.
\label{eq:timevark}
\end{equation}
As a result, the only modification to the foregoing development is that we need to implement a Dirichlet process for the coefficients for each segment $m$ in order to obtain clustering for each partition.  We assume that the Dirichlet processes are exchangeable across partitions, i.e., they share a common base measure $G$ and concentration parameter $\alpha$.

\subsection{Prior Distributions and Model Fitting}\label{sec:priors_fitting}

Here, we present model fitting for specifications presented in (\ref{eq:timevar})-(\ref{eq:ar_ou}). In our example below, we cluster on the coefficients associated with the exogenous variables.  That is, we seek to cluster with regard to the functional response to environmental variables.  So, we do not cluster on the variables capturing daily periodic behavior. As in \eqref{eq:gen}, we include an additional term $\bz_i(t) \bgamma_{i,m}^{(k)}$ in the mean to capture coefficients for which we are not interested in clustering.  These coefficients are modeled with customary vague normal priors so there are no ties in the $\bgamma_{im}^{(k)}$).

Model fitting requires prior distributions for $\bgamma_{im}^{(k)}$, $\bbeta_{im}^{(k)}$, $A_{kl}$, $\Lambda_{i}^{(k)}$, $\tau^{2(k)}$, covariance parameters ($\phi_{j}^{(k)}$) for the exponential covariance function), as well the joint DP for $\bbeta^{(k)}_{im}$. Multivariate normal prior distributions for $\bgamma_{im}^{(k)}$ are adopted, as is customary because of amenability to Gibbs sampling.

Components of the coregionalization matrix $\bA$ and $\bLambda^{(k)}$ are not uniquely identifiable without constraints. To remedy this, we constrain all components on the diagonal of $\bA$ to be 1 and all components above the diagonal to be 0. Then, lower diagonal elements are assumed \emph{a priori} to be Gaussian with high variance. Prior specifications for $\Lambda_{i}^{(k)}$ depends on the approach used \citep[e.g.][]{lopes2004,bhattacharya2011}. Our approach relies on ordering $\phi_{j}^{(k)}$ for each GP factor indexed by $j$. Thus, the factor loadings represent how much each site loads onto GPs with slower or faster temporal decays in autocorrelation. With this approach, each $\Lambda_{i}^{(k)}$ is given a weakly informative multivariate normal prior distribution. Gamma, log-normal, or uniform prior distributions for $\phi_{j}^{(k)}$ are common. Alternatively, it is common to fix $\phi_{j}^{(k)}$ for identifiability reasons \citep{zhang2004} using maximum likelihood or variogram fitting. In fact, we fixed the $\phi$'s (see below) and conducted some sensitivity analysis. Lastly, we use inverse-gamma prior distributions for $\tau^{2}_k$.


As in Section \ref{sec:dp_intro}, we specify $\bbeta_{im}^{(k)}$ through a joint Dirichlet process, where the clustering varies month-to-month. For each time window $m$, the joint prior distribution for $\bbeta_{im}^{(1)},...,\bbeta_{im}^{(K)}$ is
\begin{equation}
\pi(\bbeta_{im}^{(1)},...,\bbeta_{im}^{(K)}) = \prod^{n_s}_{i = 1} \left[ \frac{\alpha}{\alpha + i  - 1} \prod^{K}_{k = 1} \mathcal{N} \left( \bbeta_{i,m}^{k} | m^{\beta^{(k)}}, V^{\beta^{(k)}} \right) + \frac{1}{\alpha + i - 1} \sum_{l < i} \prod^{K}_{k = 1} \delta_{\bbeta_{l,m}^{(k)} } \left( \bbeta_{i,m}^{(k)}  \right) \right], \\
\end{equation}
where $\delta_{a}(x)$ is a Dirac-delta function of $x$ that has a point mass at $a$. This prior distribution assumes that the regression coefficients for all variables cluster or group together. This assumption could be relaxed to have independent Dirichlet process prior distributions for the regression coefficients for each outcome.
To complete this specification, the concentration parameter $\alpha$ could have a prior distribution with a strictly positive support or could be fixed to control clustering behavior.  We conducted sensitivity analysis using fixed values of $\alpha$.
Using the prior distributions given here, closed-form full conditional distributions are available for all parameters except $\phi_{j}^{k}$. These are presented in Appendix \ref{app:gibbs}.

\section{Analysis of Hourly Pollution Levels in Mexico City}\label{sec:mexico_city}

\subsection{Data}\label{sec:data}

\begin{figure}[H]
\begin{center}
\includegraphics[width=.4\textwidth]{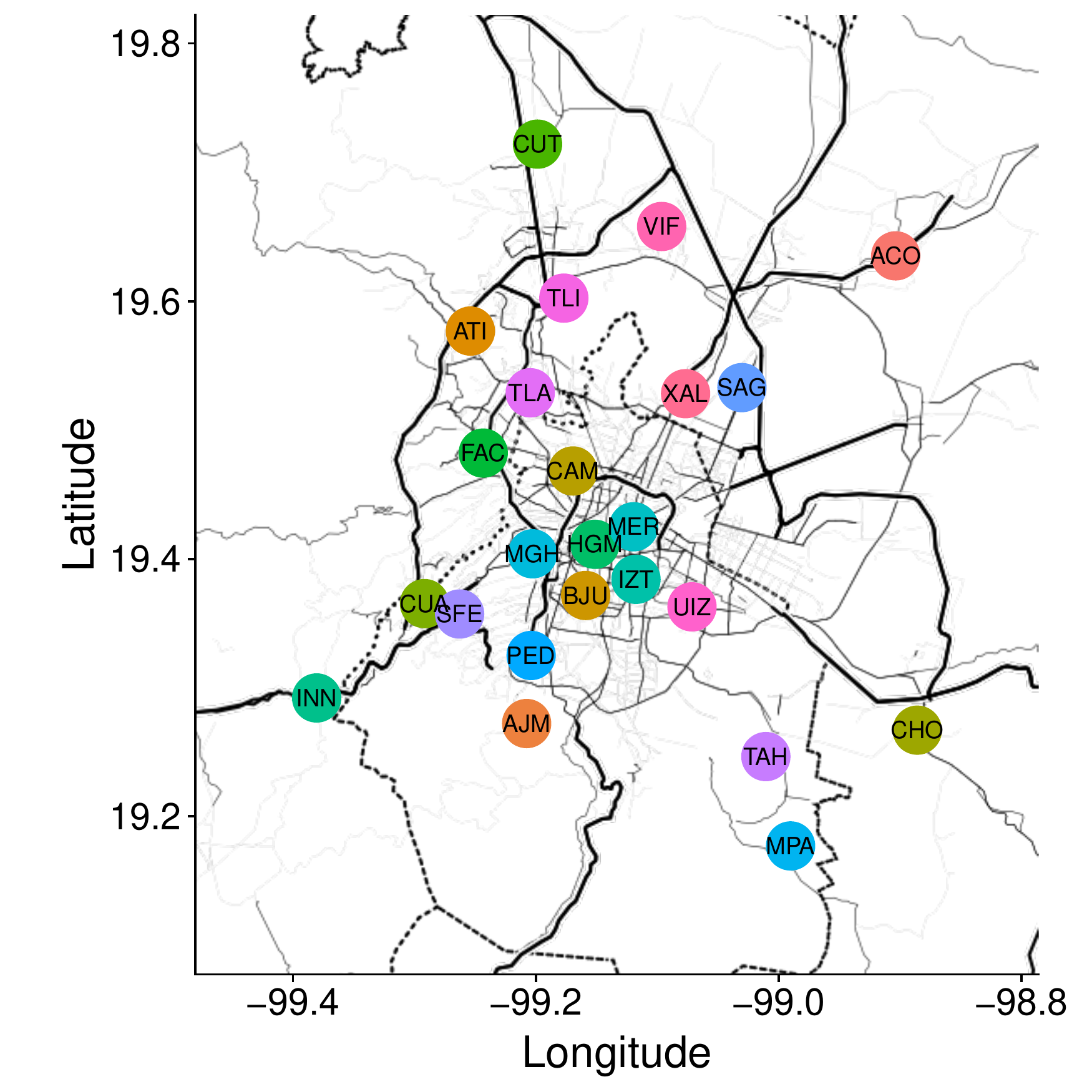}
\end{center}

\vspace{-5mm}

\caption{Station locations and abbreviated names}\label{fig:stat_loc}
\end{figure}

Here, we consider hourly ozone and \pmt measurements from 24 stations in Mexico City during 2017. The locations of the 24 stations and their abbreviated names are plotted in Figure \ref{fig:stat_loc}. The data are available under Datos/Horarios/Contaminante at \url{http://www.aire.cdmx.gob.mx/default.php}. In total, each station has 8760 ozone and \pmt measurements.\footnote{Missing hourly measurements were imputed using the corresponding measurements at the nearest station within the same region. If no stations in that region recorded a measurement at that time, then the nearest station in a different region provided the missing value. This was done prior to our receiving the data for analysis.}

\begin{figure}[H]
\begin{center}
\includegraphics[width=\textwidth]{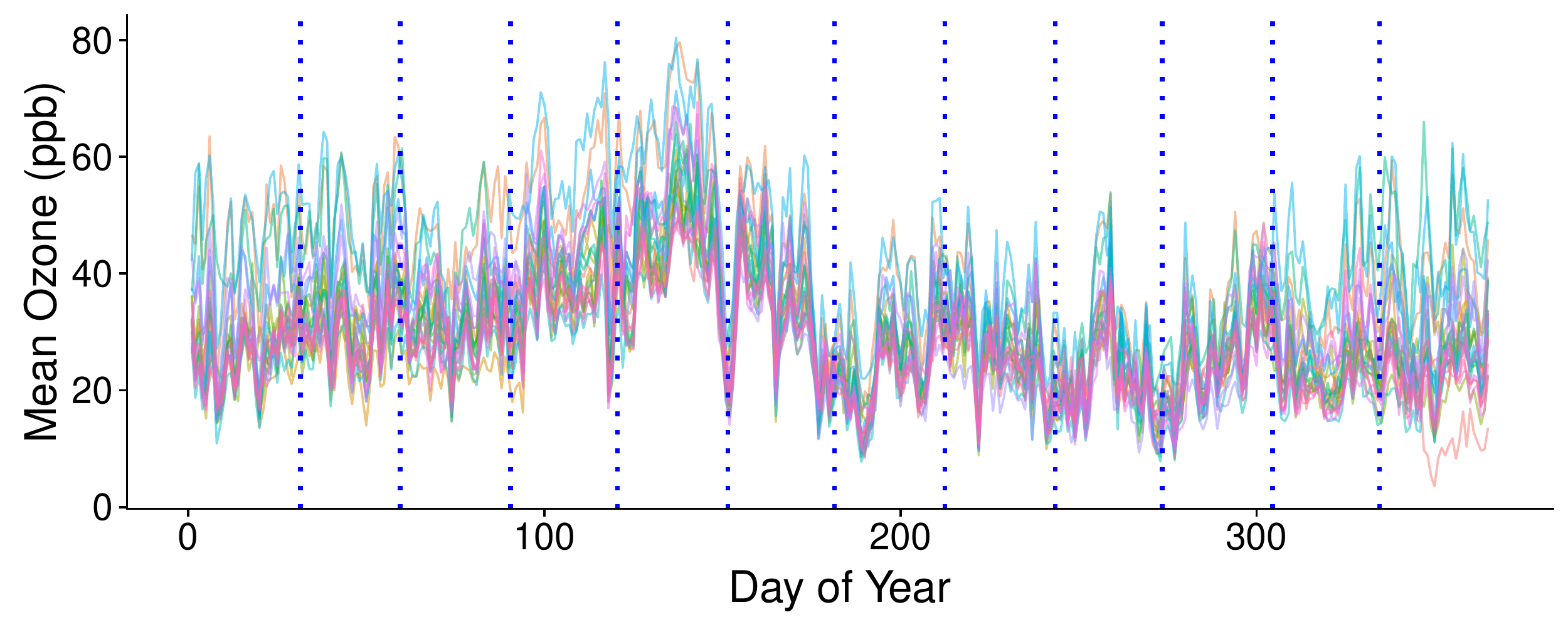}
\includegraphics[width=\textwidth]{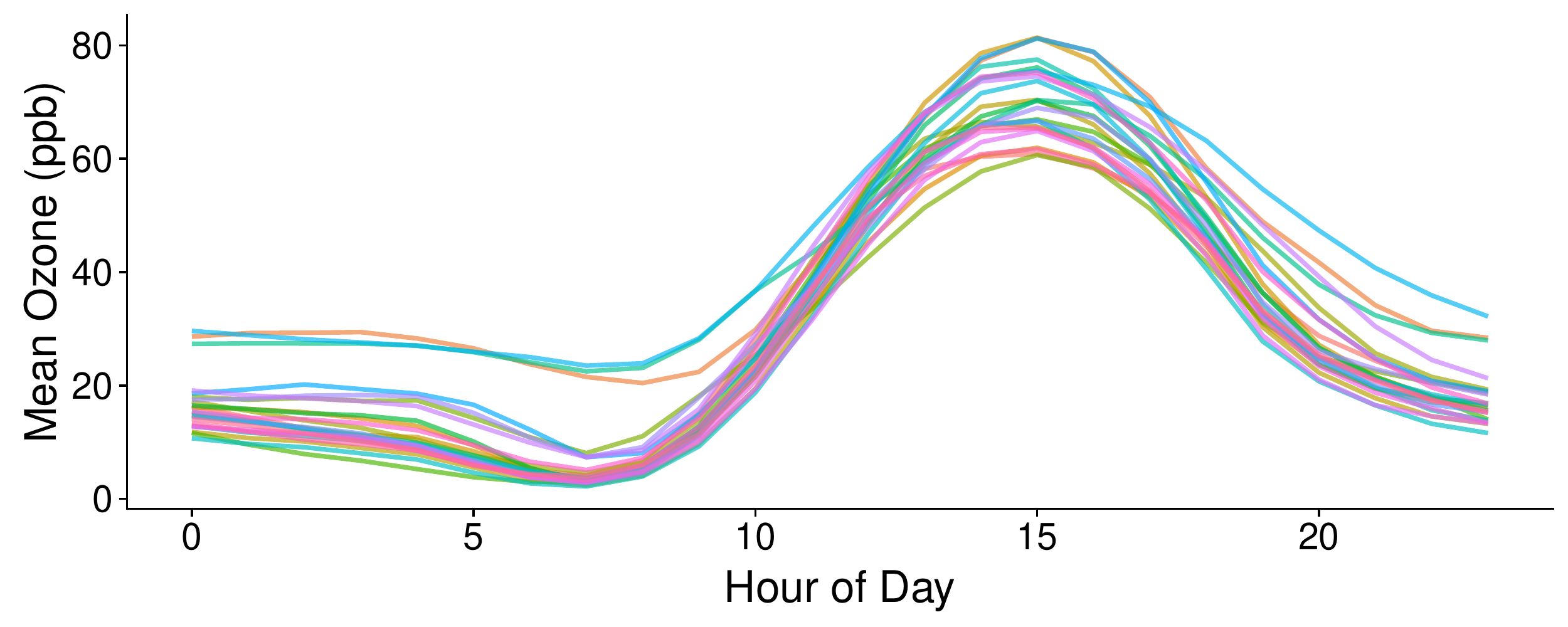}
\includegraphics[width=0.45\textwidth]{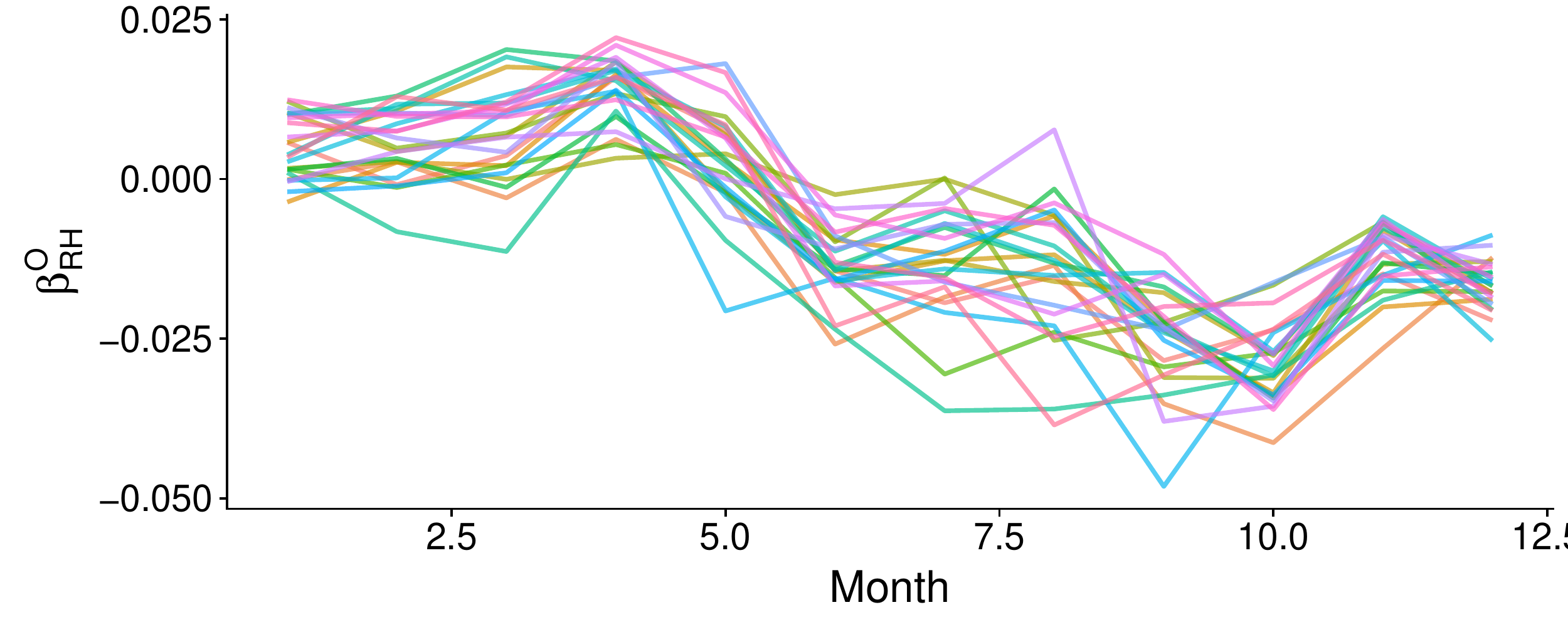}
\includegraphics[width=0.45\textwidth]{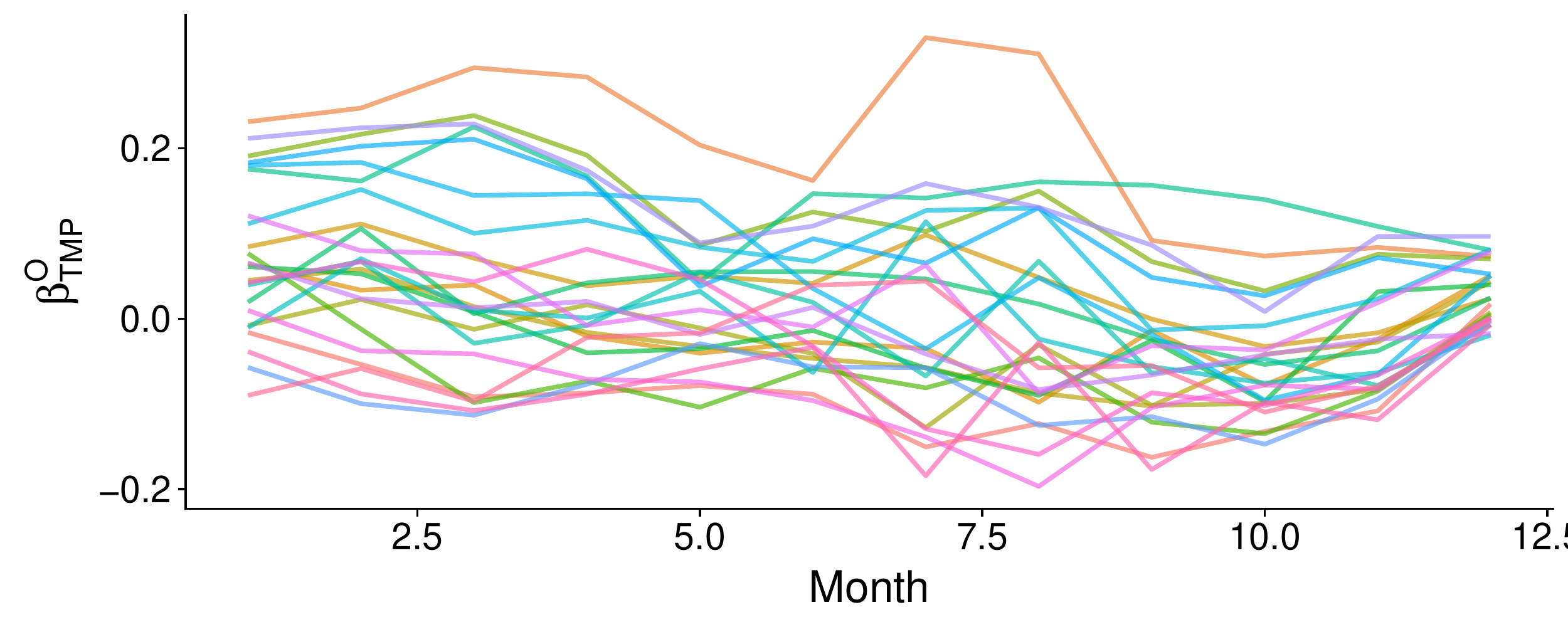}
\includegraphics[width=0.45\textwidth]{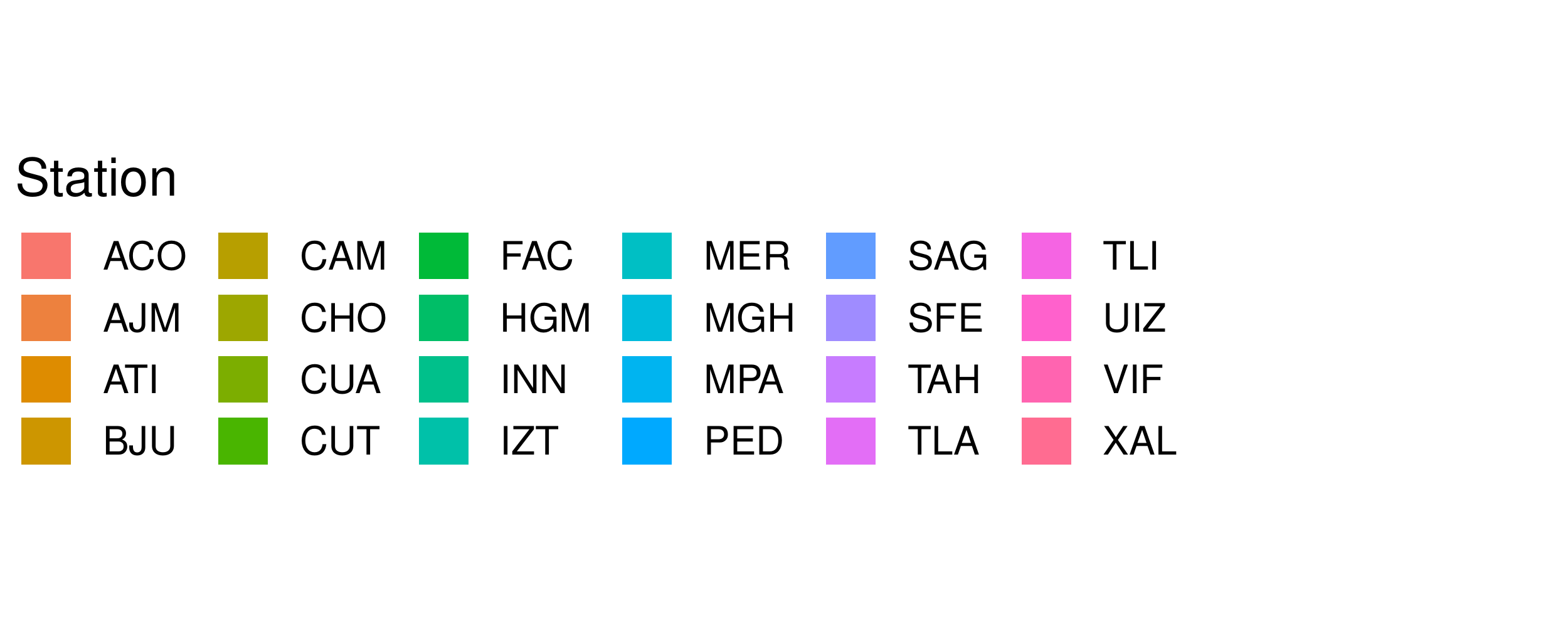}

\end{center}
\caption{Summaries of station-specific ozone levels and station-specific time-varying regression models. (Top) Daily averages of ozone levels (Top-middle) Hourly average. (Bottom-left) Monthly relative humidity regression coefficients (Bottom-right)  Monthly temperature regression coefficients  }\label{fig:ozone_summaries}
\end{figure}

\begin{figure}[H]
\begin{center}
\includegraphics[width=\textwidth]{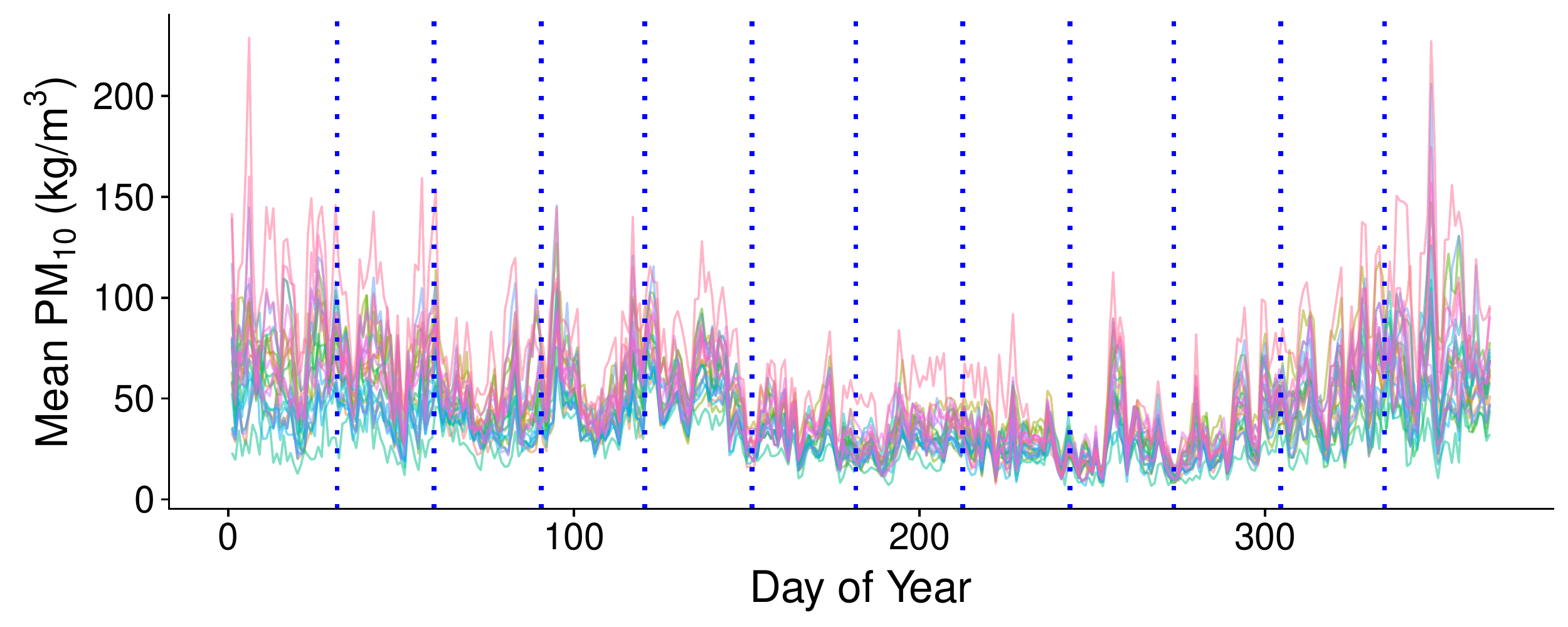}
\includegraphics[width=\textwidth]{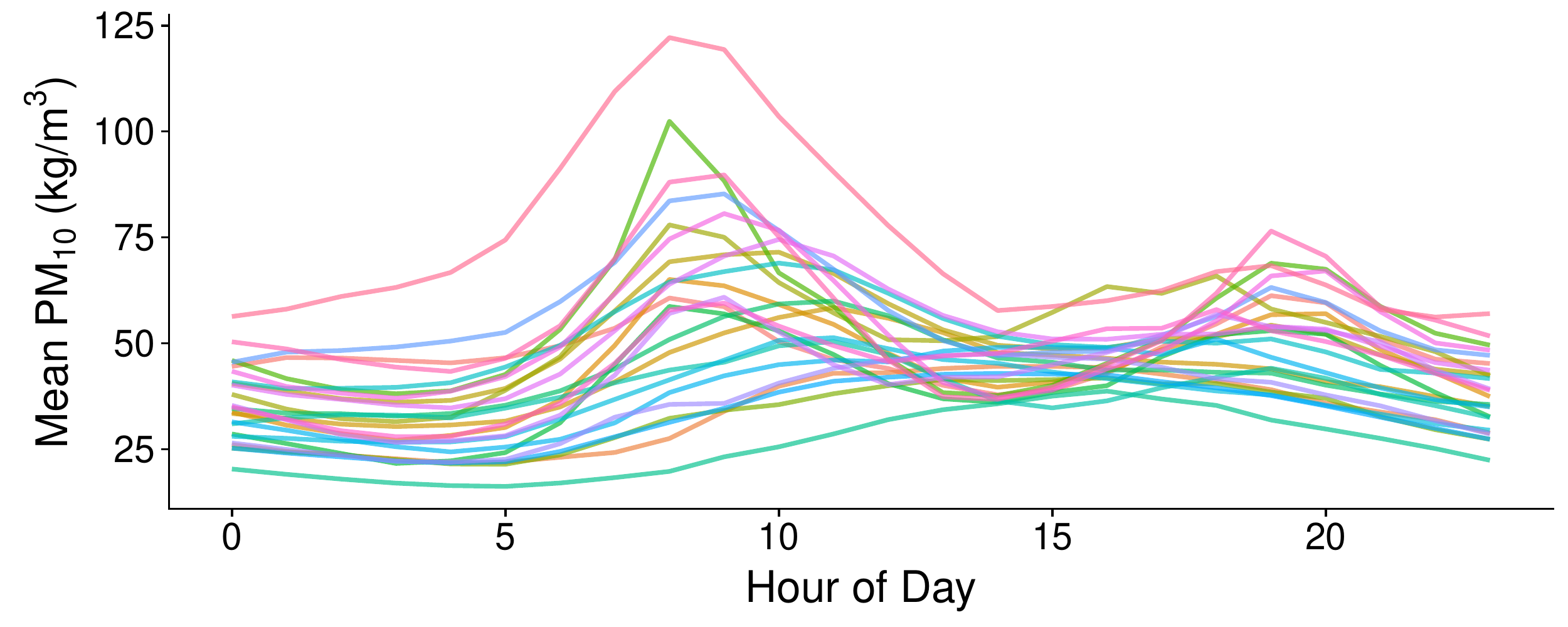}
\includegraphics[width=0.45\textwidth]{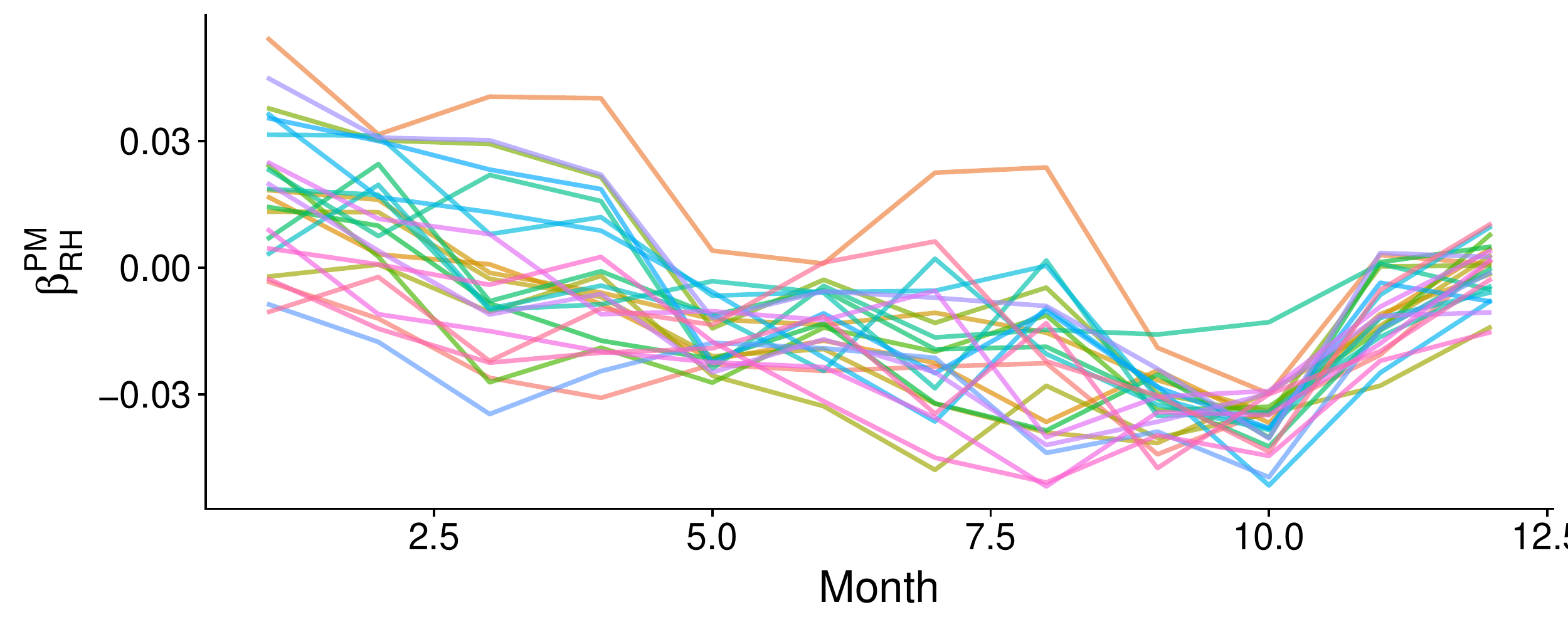}
\includegraphics[width=0.45\textwidth]{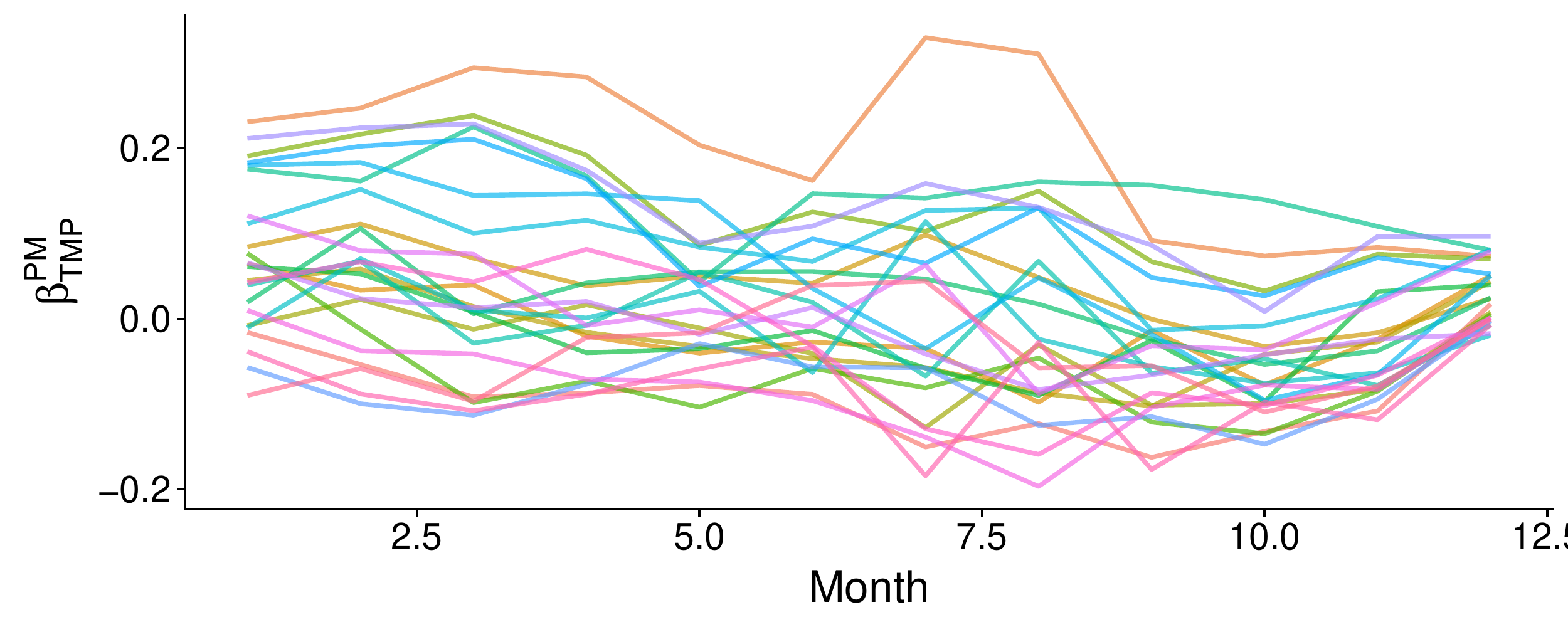}
\includegraphics[width=0.45\textwidth]{averages_legend}

\end{center}
\caption{Summaries of station-specific \pmt levels and station-specific time-varying regression models. (Top) Daily averages of ozone levels (Top-middle) Hourly average. (Bottom-left) Monthly relative humidity regression coefficients (Bottom-right)  Monthly temperature regression coefficients }\label{fig:pm10_summaries}
\end{figure}

For both ozone and \pmt, we examine patterns in the data that motivate modeling decisions. For each station, we plot the average of each day, averaged over all hours in the day, and the hourly average, averaged over all hours of the day. In addition, we examine how the relationships between covariates (relative humidity and temperature) and outcomes (ozone and \pmt) vary month-by-month for each station. To explore this, we fit a linear model for both outcomes at each station for each month using relative humidity and temperature as covariates. Exploring these patterns allows us to assess how relationships between covariates and outcomes differ station-to-station and how they vary over 2017. These patterns for ozone are given in Figure \ref{fig:ozone_summaries} and for \pmt in Figure \ref{fig:pm10_summaries}.

\vspace{-2mm}

Figure \ref{fig:ozone_summaries} shows significant variability in ozone levels over the year and as a function of time-of-day. Ozone levels peak in the Spring (April and May are the peak ozone season). As a function of the time-of-day, ozone levels are highest in the early afternoon, dip in the evening, and reach a minimum in the morning. While these trends are common across stations, there is variability in ozone levels and patterns between stations. In addition to the variability in ozone levels, the results from the monthly regression models show significant variability in the intercept and regression coefficients for humidity and temperature month-to-month, suggesting potential benefit of time-varying effects. However, regression coefficients are negative for relative humidity and positive for temperature.

The annual trends in \pmt are less clear than they are for ozone; however, the highest levels in \pmt occur in the winter. The hourly averages of \pmt in Figure \ref{fig:pm10_summaries} exhibit clear peaks around 8 AM and 7 or 8 PM. Like the regression coefficients for ozone, the effects of humidity and temperature on \pmt vary greatly over 2017 and station-to-station. Again, these summaries indicate that a hierarchical model with dynamic regression coefficients may be appropriate.

\subsection{Model}\label{sec:mexico_city_model}


For the Mexico City data, let $Y^{O}_i(t)$ be ozone levels in parts per billion (ppb), $Y^{PM}_i(t)$ be \pmt concentration in $kg/m^3$  at location $i$ and hour $t$. In addition, let $\bx_i(t)$ be temperature in degrees Celsius at station $i$ and hour $t$. We model ozone on the square root scale, following \cite{gelfand2003,sahu2007,berrocal2010,white2018a,white2018b}. We model \pmt on the log scale as in \cite{cocchi2007,huang2018,white2018a}. In addition, we center and scale the transformed outcomes so that the likelihoods for each outcomes are similar. In both cases, these transformations stabilize variance between model mean and variance of residuals and improve predictive performance. We induce clustering hierarchically through an exchangeable monthly Dirichlet process on the  regression coefficients. We index months by $m$ such that the models in \eqref{eq:ozone} and \eqref{eq:pm10} apply to $t \in m$.

Explicitly, we have
\begin{align}
\sqrt{ Y^{O}_i(t)} &= \bz(t)^T \bgamma^O_{im} +  \bx_i(t)^T \bbeta_{im}^{O} + \eta_i^O(t) + \epsilon_i^O(t) \label{eq:ozone} \\
\log\left(Y^{PM}_i(t) \right) &= \bz(t)^T\bgamma^{PM}_{im} + \bx_i(t)^T \bbeta_{im}^{PM} + \eta_i^{PM}(t) + \epsilon_i^{PM}(t) \label{eq:pm10},
\end{align}
where $\epsilon_i^{O}(t)$ and $\epsilon_i^{PM}(t)$ are assumed to be Gaussian with a mean of 0 and variance of $\tau^2_1$ and $\tau^2_2$, respectively.
Here, we let coefficients ($\bgamma^O_{im}$ and $\bgamma^{PM}_{im}$) for $\bz(t)^T$ (an intercept and sine and cosine terms with periods of 24 hours) be at monthly scale.
As noted above, we do not include these parameters as a part of the Dirichlet process because the goal is clustering stations on with regard to response to meteorological variables (humidity and temperature). 
We fix the number of factors to $r = 3$. For each factor, we fix the decay parameters $\phi^{(k)}_j$ such that $\phi^{(k)}_1= 1/24$, $\phi^{(k)}_2 = 1/3$, and $\phi^{(k)}_3 = 1$ for all $k$.  The first factor corresponds to long-range autocorrelation (with an effective range of 72 hours), the second factor captures middle-range autocorrelation (an effective range of nine hours), and the third factor accounts for short-term autocorrelation (an effective range of three hours).

We fit our model using the Gibbs sampler presented in Section \ref{sec:methods_models} and Appendix \ref{app:gibbs}. Specifically, we ran the model for 150,000 iterations, discarding the first 50,000 samples. Then, we thinned the remaining 100,000 sample to 10,000 samples, keeping every 10th sample.

In Appendix \ref{app:sensitivity}, we provide a discussion of the sensitivity of the model. To summarize this discussion, the number of clusters in the model is somewhat sensitive to selection of $\alpha$, specifications of the base measure of the Dirichlet process prior, and prior distributions on $\tau^2_k$. In addition, we found that joint clustering was driven exclusively by ozone. If we used independent Dirichlet processes for ozone and \pmt, then clustering was limited to ozone.

\subsection{Results}\label{sec:results}

\begin{figure}[H]
\begin{center}
\includegraphics[width=0.45\textwidth]{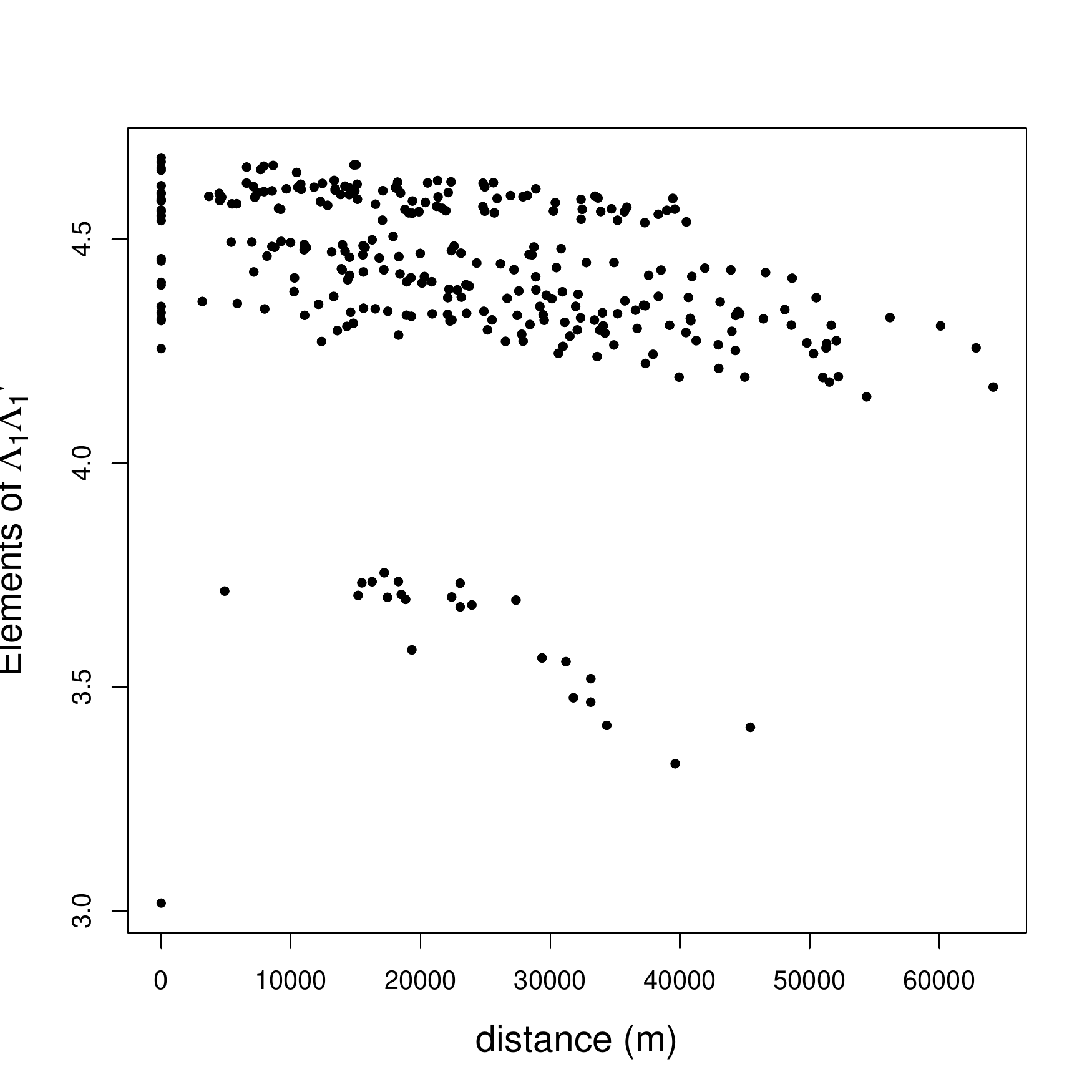}
\includegraphics[width=0.45\textwidth]{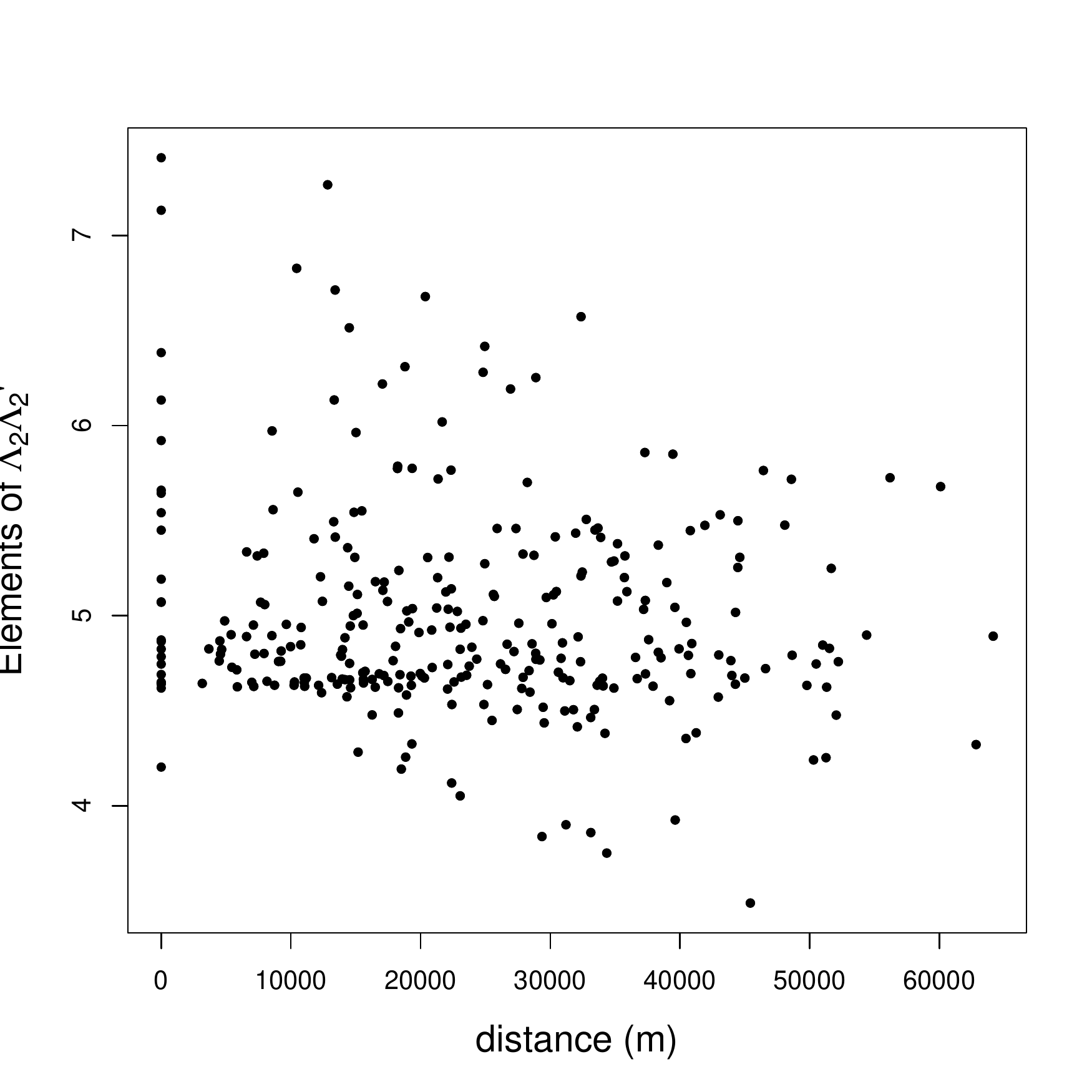}
\end{center}
\caption{ Mean elements of $\Lambda^{(k)}{\Lambda^{(k)}}^T$ plotted against distance between sites. Covariance patterns for (Left) ozone (right) \pmt. }\label{fig:distance_LtL}
\end{figure}

\begin{figure}[H]
\begin{center}
\includegraphics[width=0.45\textwidth]{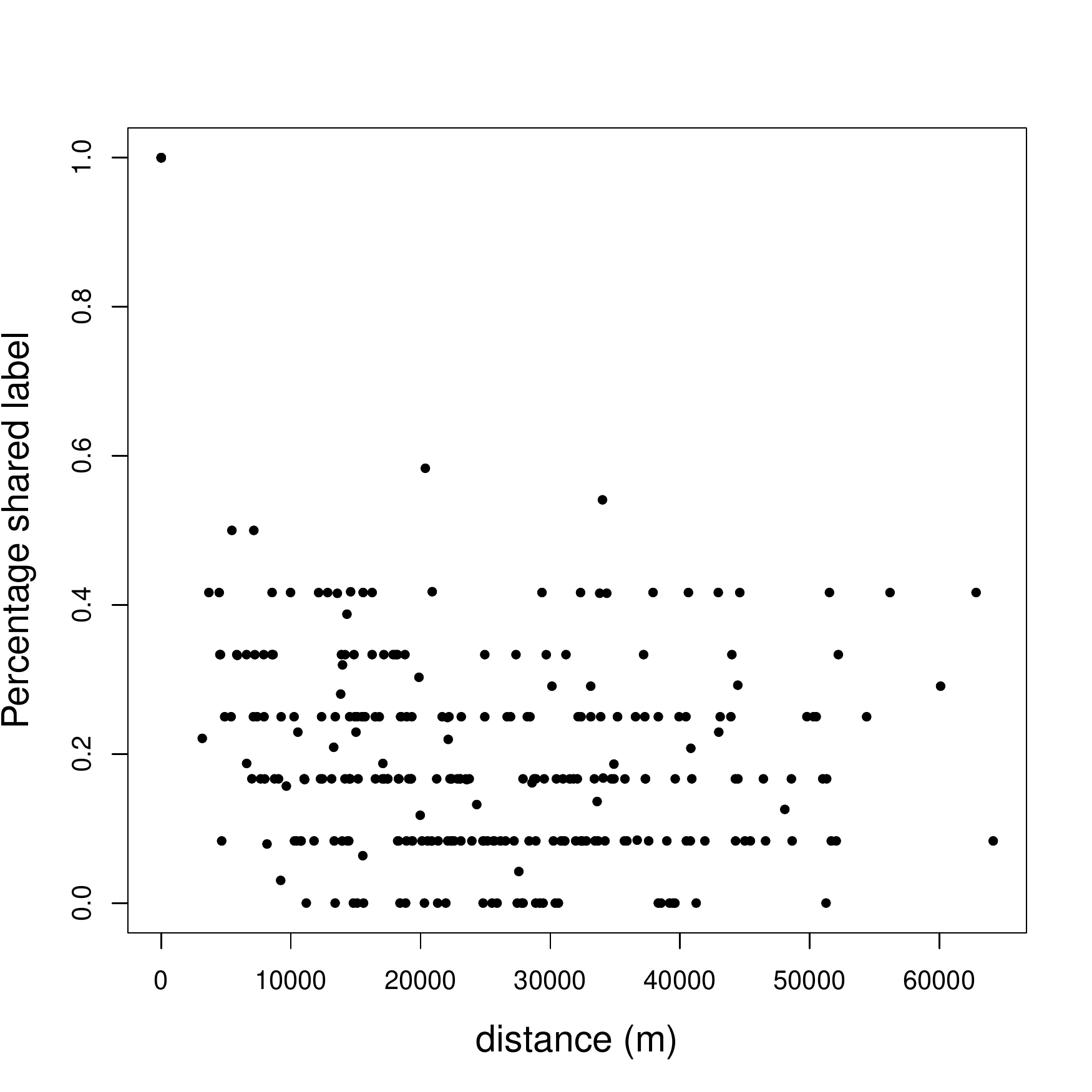}
\end{center}
\caption{Proportion of times sites share a functional label, averaged over months and iterations. Shared labels indicates shared regression coefficients.}\label{fig:distance_shared}
\end{figure}

Our primary goal is to explore how the clustering of stations evolves over time.  However, we first address possible issues with regard to our modeling choices. To examine the benefit of modeling ozone and \pmt jointly, we note the significance of the coregionalization parameter $a_{12}$; it has a posterior mean of $0.158$ with 95\% credible interval $(0.141,0.180)$. To address not including space in our model, because $\Lambda^{(k)}{\Lambda^{(k)}}^T$ represents the between-site covariance in our model, we plot the components of these matrices against the distance between sites in Figure \ref{fig:distance_LtL}.  We can assess whether the factor loadings are capturing spatial behavior. If there was consequential spatial dependence, we would expect a negative relationship between the between-site covariance and distance. However, the plots in Figure \ref{fig:distance_LtL} do not suggest any strong spatial patterns. We also plot the proportion of times that sites share a label, averaged over months, against the distance between sites in Figure \ref{fig:distance_shared}. Again, there does not appear to be a very strong relationship between clustering and space.

\begin{figure}[H]
\begin{center}
\includegraphics[width=0.45\textwidth]{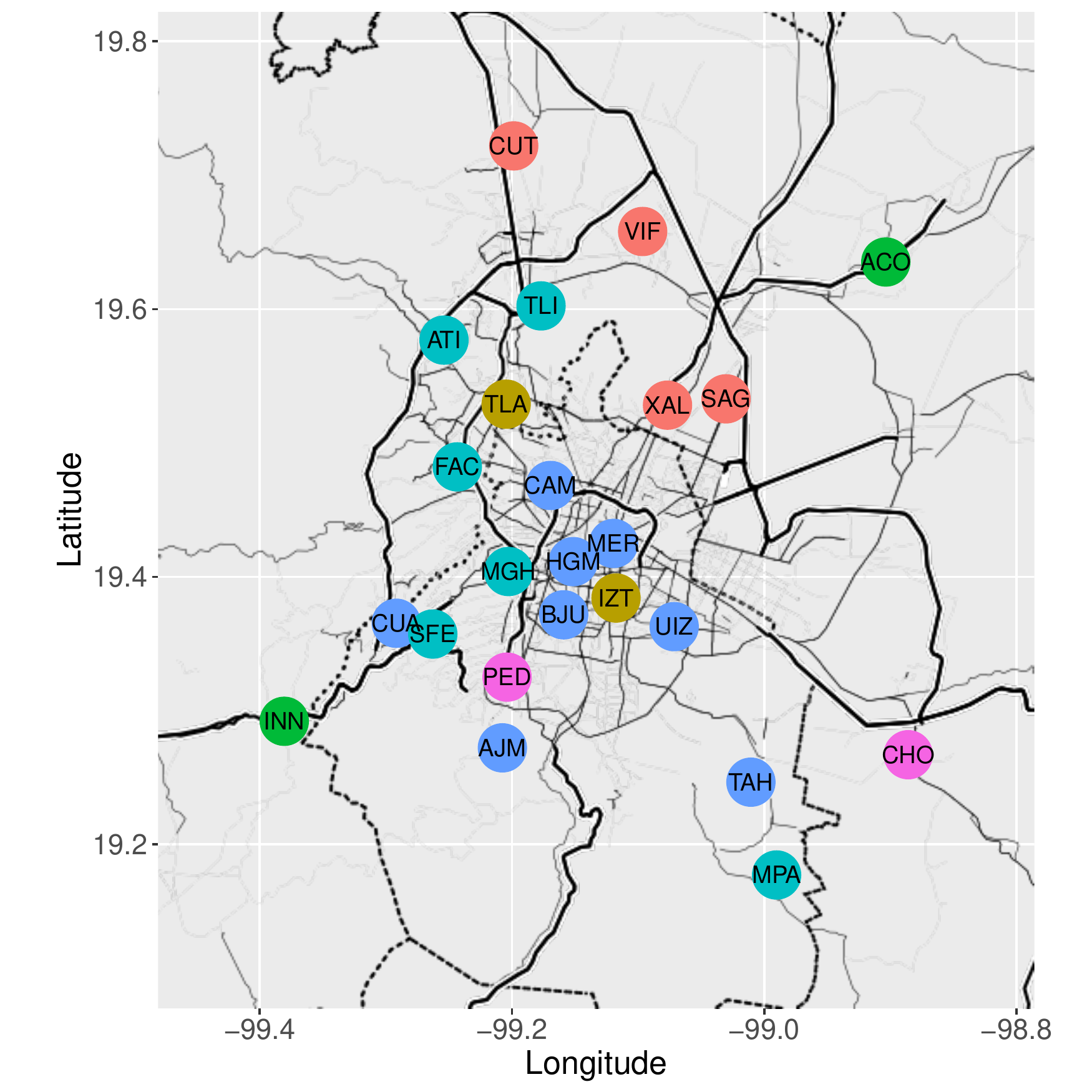}
\includegraphics[width=0.45\textwidth]{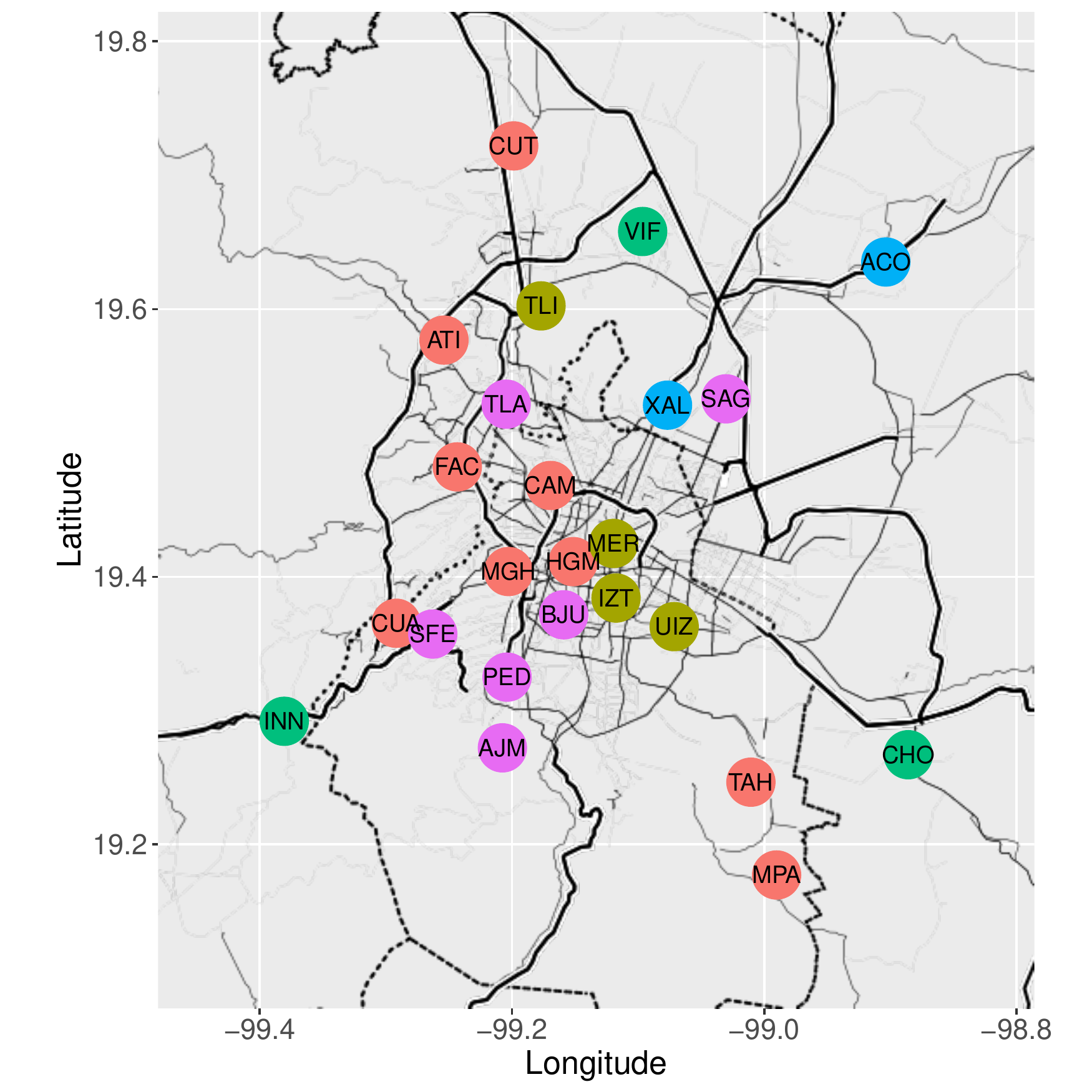}
\end{center}
\caption{Cluster labels indexed by color for (Left) May and (Right) December. }\label{fig:loc_clust}
\end{figure}

Although Figures \ref{fig:distance_LtL} and \ref{fig:distance_shared} suggest that the data do not have a strong spatial component after accounting for temporal autocorrelation, looking at clustering for individual months, we find some spatial patterns. For illustration, we provide the station location and cluster label (indexed by color) for May and December in Figure \ref{fig:loc_clust}. In May, there appears to be a clear cluster in the north (in red) and three clusters in central Mexico City (teal, blue, and gold). In December, we see three clusters with strong spatial patters: red is concentrated in the east, purple in the southeast, and gold in central Mexico City. Within each month, these maps seem to indicate that the clustered regression coefficients capture spatial similarity that our Gaussian process model does not.

\begin{figure}[H]
\begin{center}
\includegraphics[width=0.45\textwidth]{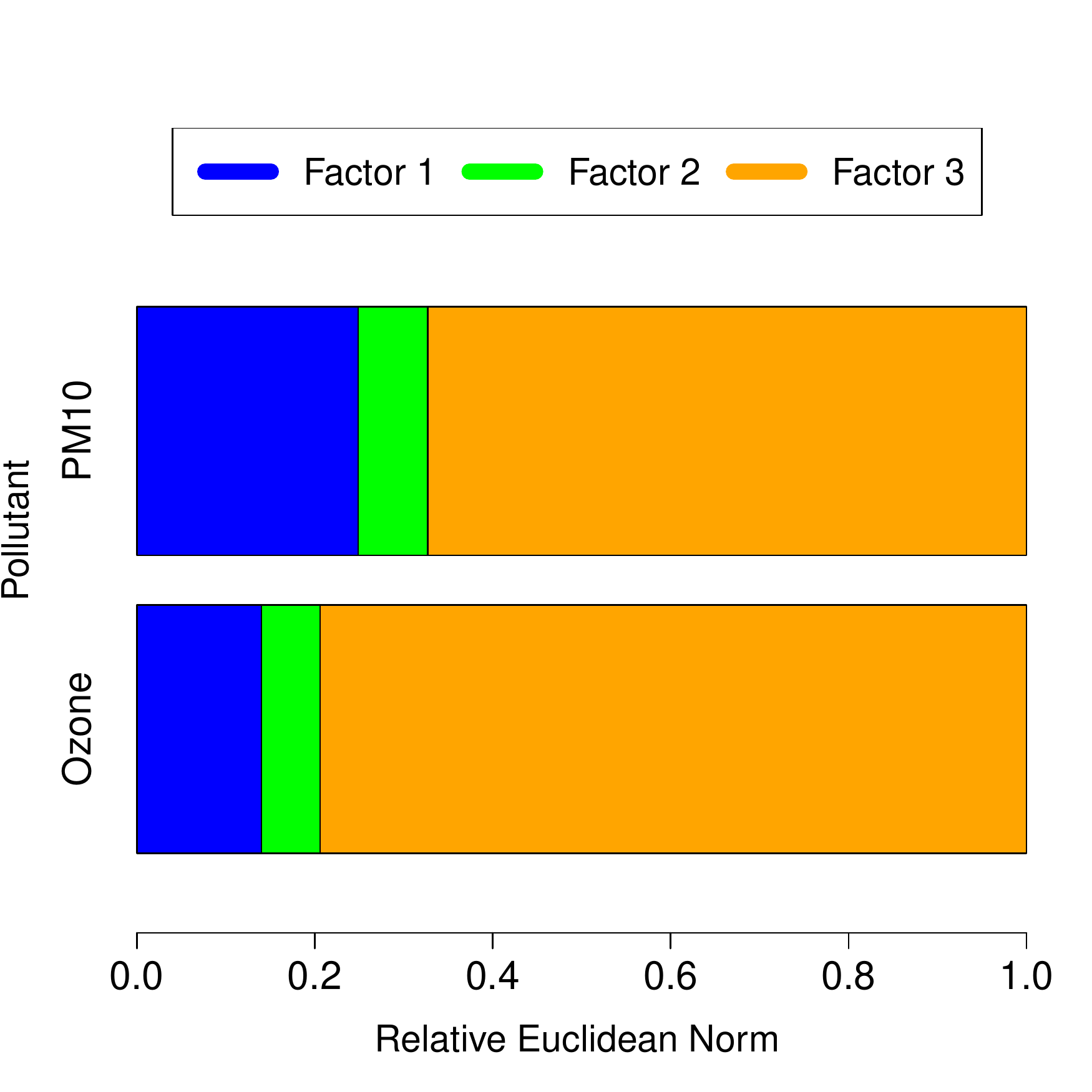}
\end{center}
\caption{Relative Euclidean norms of the columns of $\bLambda^{(k)}$.}\label{fig:lam_norm}
\end{figure}

Again, we use $3$ GP factors, with decay parameters fixed, as above.  To examine the relative importance of these factors, we compute and plot the relative Euclidean norms of the columns of $\bLambda^{(k)}$ in Figure \ref{fig:lam_norm}. Interestingly, the long and short range factors account for 93.4\% and 92.2\% of the factor model for ozone and \pmt, respectively. Therefore, three factors appear to be sufficient in this application.

\begin{figure}[H]
\begin{center}
\includegraphics[width=0.8\textwidth]{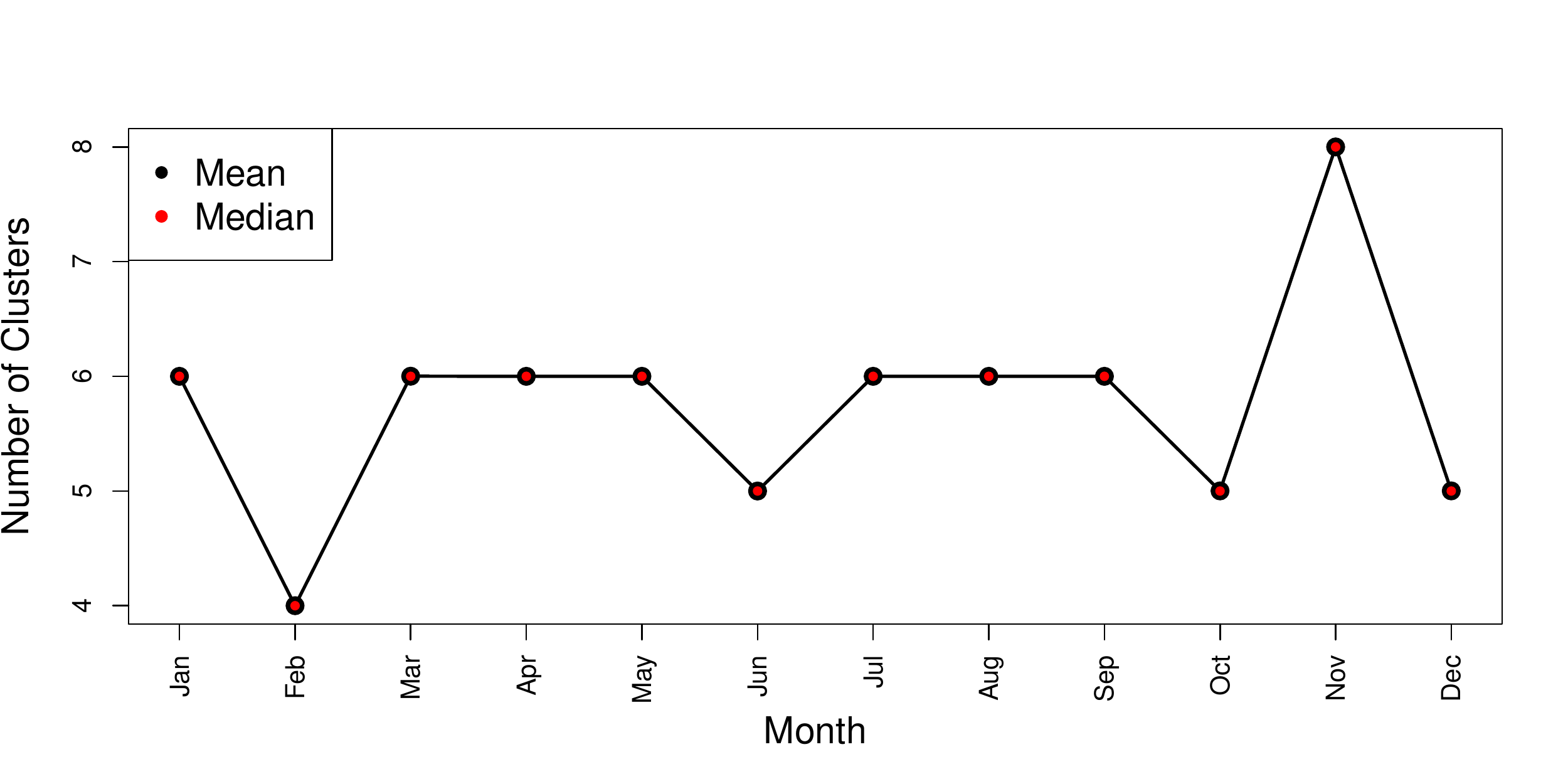}
\end{center}
\caption{ Number of clusters for each month.}\label{fig:clust_num}
\end{figure}

In this analysis, the mean number of clusters varies somewhat month-to-month. To illustrate this, we plot the number of clusters as a function of month in Figure \ref{fig:clust_num}. Averaging over all months and iterations of our sampler, the mean number of clusters is 5.75.  We note that clustering between sites is driven by ozone in this example. In fact, when we use independent DPs for ozone and \pmt, there is no clustering for \pmt. We also found that the number of clusters was relatively robust to the prior distribution on error terms $\tau^2_k$. However, we did see that the number of clusters was somewhat sensitive to the selection of the concentration parameter $\alpha$ and the base measure for regression coefficients. We discuss this in more detail in Appendix \ref{app:sensitivity}.

\begin{figure}[H]
\begin{center}
\includegraphics[width=0.4\textwidth]{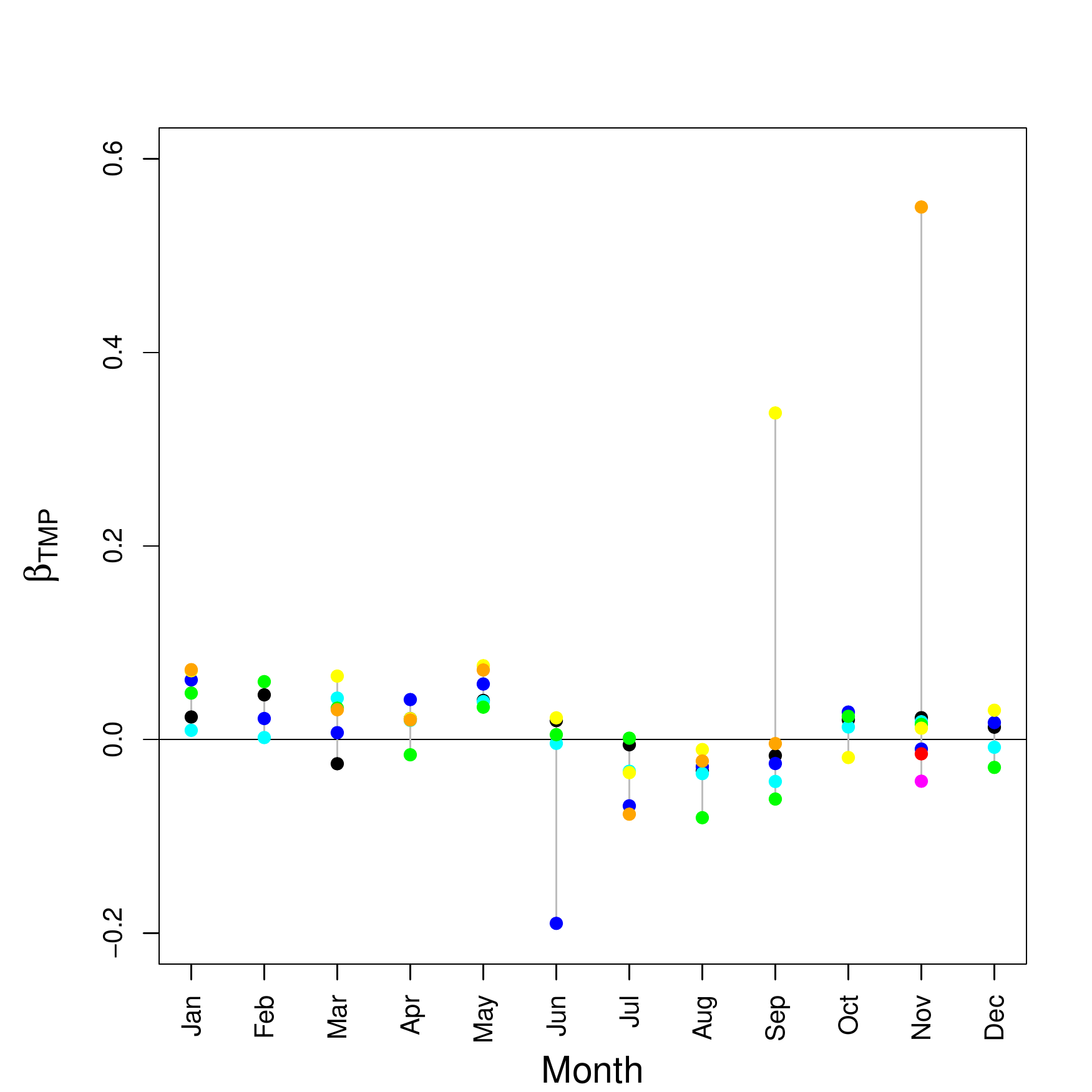}
\includegraphics[width=0.4\textwidth]{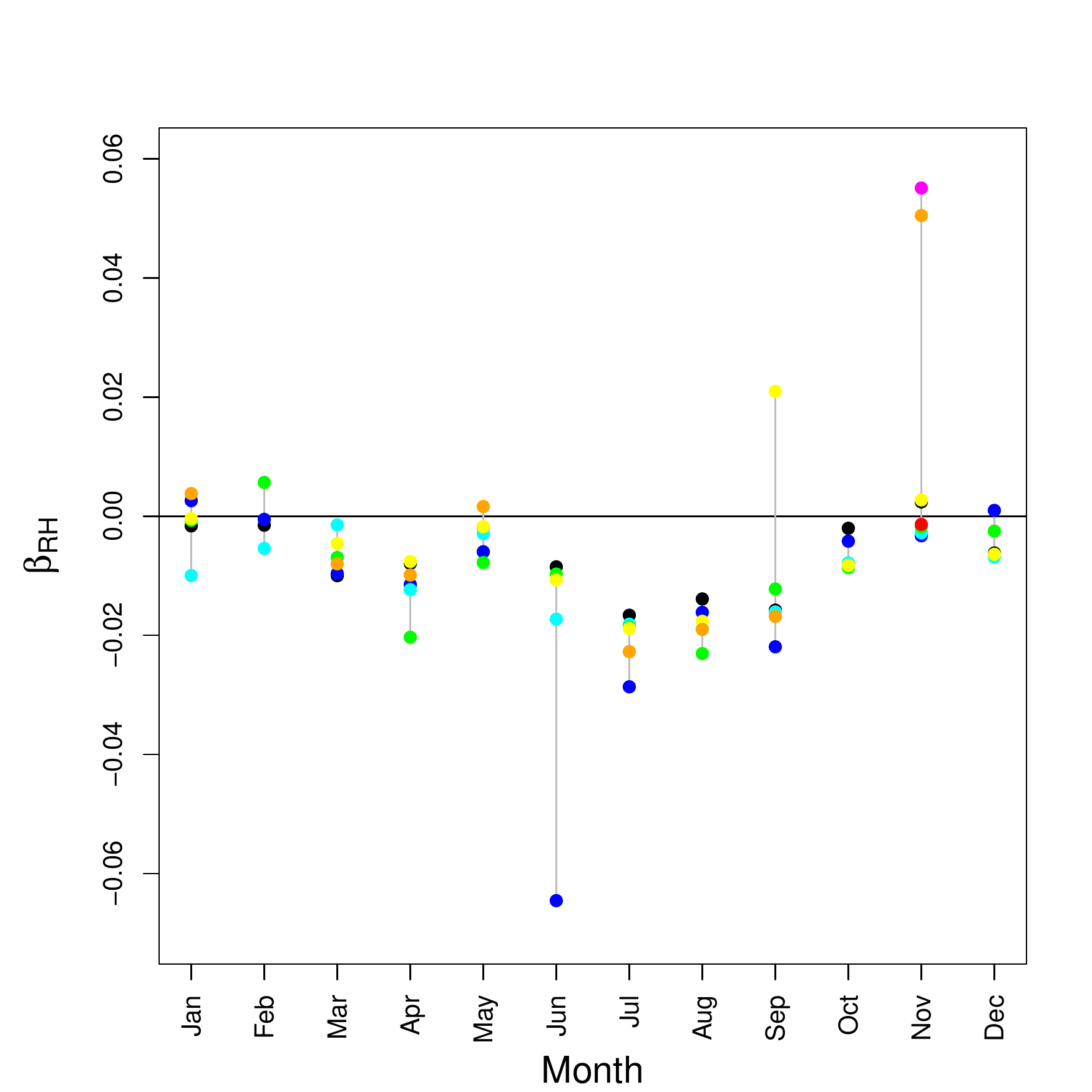}
\includegraphics[width=0.4\textwidth]{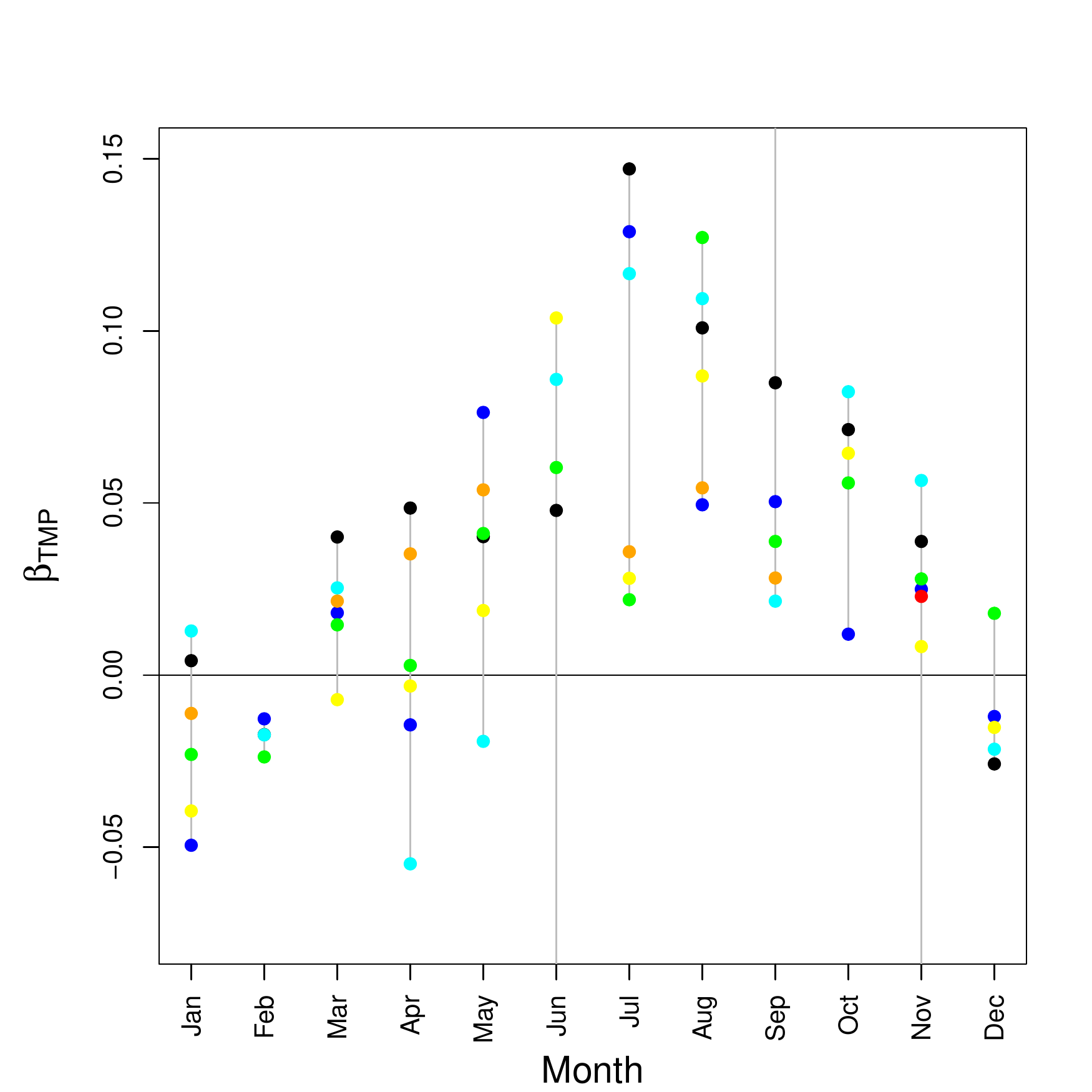}
\includegraphics[width=0.4\textwidth]{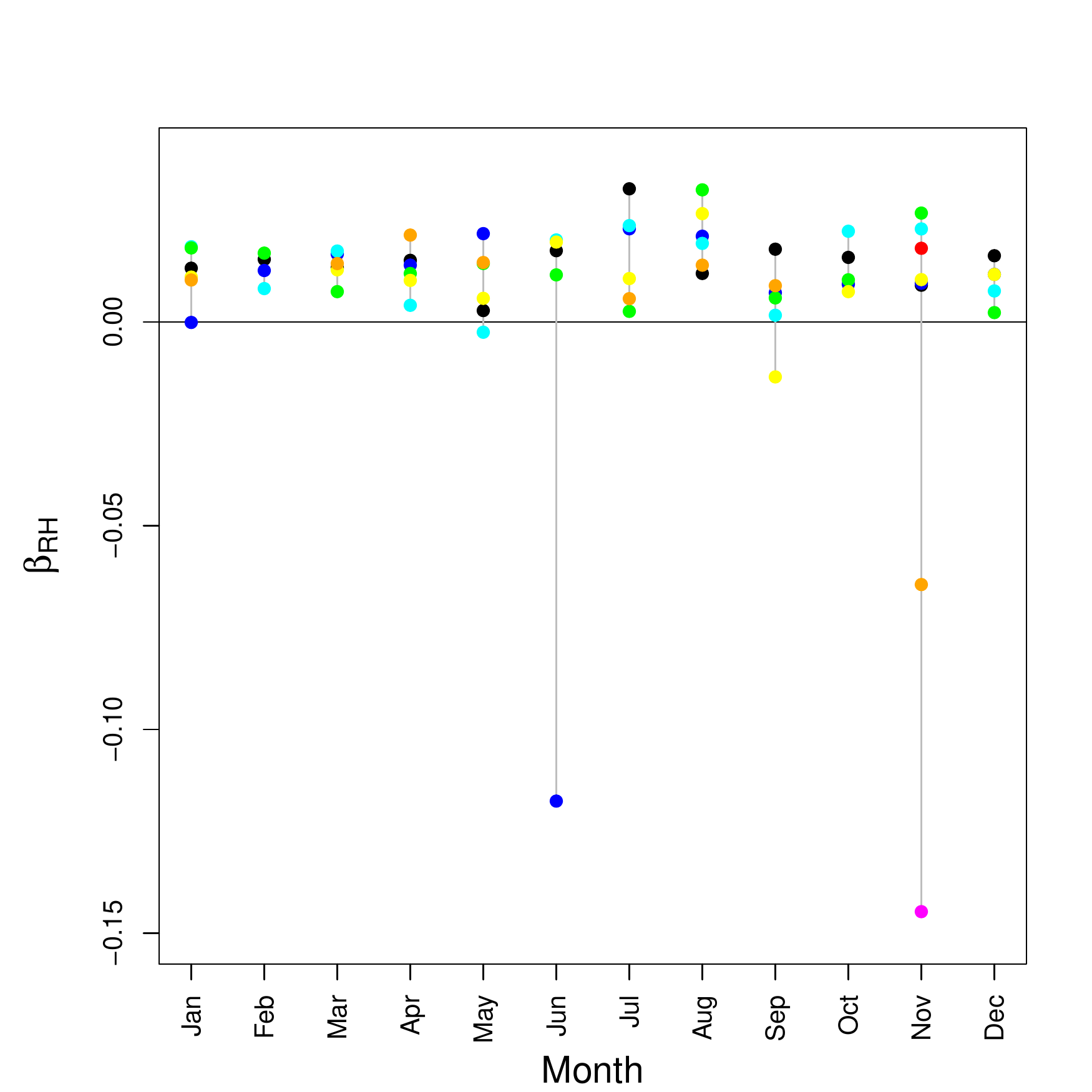}
\end{center}
\caption{ Posterior means for regression coefficients for ozone and \pmt plotted by month. The estimated effect on ozone of (Top-Left) temperature and (Top-Right) relative humidity. The estimated effect on \pmt of (Bottom-Left) temperature and (Bottom-Right) relative humidity. For each month, the same dot color in these two plots indicated the same cluster.  }\label{fig:ref_coef}
\end{figure}

Finally, we examine the regression coefficients for ozone and \pmt. These should be interpreted in the context of our multivariate GP model with time-varying intercepts and seasonal terms. In Figure \ref{fig:ref_coef}, we plot the cluster-specific posterior means for the regression coefficients for the effect of temperature and relative humidity on temperature. These show significant variability over the course of the year and within the month (i.e., between clusters). For January to May, temperature has a positive relationship with ozone for most clusters, but there is significant variability during the rest of the year. Humidity appears to have a negative relationship for most clusters from March to October. Temperature and \pmt generally have a positive relationship from March to November. Relative humidity has a positive relationship with \pmt during most of the year.

\section{Conclusions and Future Work}\label{sec:conclusions}

In this paper, we have introduced a modeling framework for multivariate functional data over time and across sites with the goal of time-varying clustering on regression coefficients. We presented a continuous time scalable multivariate GP model that uses factor analysis to capture dependence between sites and coregionalization to capture dependence within sites. leverages a factor model to capture inter-site, co-regionalization to model covariance between variables.  We used the exponential correlation function for temporal autocorrelation, resulting in a continuous time first-order Markov process. We used an exchangeable time-varying Dirichlet process prior distributions for regression coefficients to cluster sites that have a similar joint response to exogenous variables (covariates). The resulting hierarchical model was fitted within a Bayesian framework whence we presented appropriate prior distribution and associated model fitting details. Lastly, we analyzed our motivating dataset: hourly ozone and \pmt data from Mexico City in 2017. 

Turning to future work, we could allow component specific partitions.  In fact, within the Dirichlet process mixing framework, we could attempt atom, i.e., label selection in continuous time following ideas in \cite{petrone2009}.  In a second direction, rather than assuming that DPs were exchangeable month-to-month, we could incorporate dependence between base measures through a dynamic linear model \citep{west1997}. As a third possibility, though spatial autocorrelation was not a significant component in our motivating example, in some datasets sites may exhibit spatial autocorrelation after accounting for covariates and temporal autocorrelation. In these cases, we could incorporate spatial autocorrelation between factor loadings \citep[see][for early discussion on this]{christensen2002,hogan2004}. 

\appendix

\section{Distributional Properties}\label{app:dist}

First, we derive the result in \eqref{eq:dep}.  Following the notation in Section 2.1, we seek $\text{cov}(\eta_{i}^{k}(t), \eta_{i'}^{k'}(t')$ for say $t' \geq t$.  We have

\begin{equation}
\text{cov}(\eta_{i}^{k}(t), \eta_{i'}^{k'}(t')) = \text{cov}(\sum_{j=1}^{K}A_{kj} \phi_{ij}^{(k)}(t), \sum_{j=1}^{K}A_{k'j} \phi_{i'j}^{(k')}(t'))=\\
\sum_{j=1}^{K} \sum_{j'=1}^{K} A_{kj}A_{k'j'} \text{cov}(\phi_{ij}^{(k)}(t), \phi_{i'j'}^{(k')}(t'))
\end{equation}
where $\phi_{ij}^{(k)}(t) = \sum_{l=1}^{r} \Delta_{il}^{(k)} \nu_{l}^{(j)}(t)$ and $\phi_{i'j'}^{(k')}(t') = \sum_{l=1}^{r} \Delta_{i'l}^{(k')} \nu_{l}^{(j')}(t')$.  Since the $\nu_{l}^{(j)}(t)$ are independent across $l$ and $j$, $\text{cov}(\phi_{ij}^{(k)}(t), \phi_{i'j'}^{(k')}(t'))= 0$ if $l \neq l'$ or $j \neq j'$.  However, with $j = j'$, we have $\text{cov}(\phi_{ij}^{(k)}(t), \phi_{i'j}^{(k')}(t') = \sum_{l=1}^{r} \Delta_{il}^{(k)} \Delta_{i'l}^{(k')} e^{-\phi_{(j)}(t' -t)}$.

So, $\sum_{j=1}^{K} \sum_{j'=1}^{K} A_{kj}A_{k'j'} \text{cov}(\phi_{ij}^{(k)}(t), \phi_{i'j'}^{(k')}(t') = \sum_{j=1}^{K} A_{kj} A_{k'j} \sum_{l=1}^{r} \Delta_{il}^{(k)} \Delta_{i'l}^{(k')}e^{-\phi_{(j)}(t' -t)}$.  After a little algebra this becomes $(\bLam_{i}^{(k)})^{T}\bLam_{i'}^{(k')}\sum_{j=1}^{K}A_{kj}A_{k'j}e^{-\phi_{(j)}(t'-t)}$.

Special cases of \eqref{eq:dep} used below are the following:

(i) when $t' =t$, $\text{cov}(\eta_{i}^{k}(t), \eta_{i'}^{k'}(t') = (\bLam_{i}^{(k)})^{T}\bLam_{i'}^{(k')}\sum_{j=1}^{K}A_{kj}A_{k'j} $,

(ii) when $i'=i$, $\text{cov}(\eta_{i}^{k}(t), \eta_{i'}^{k'}(t') = (\bLam_{i}^{(k)})^{T}\bLam_{i}^{(k')}\sum_{j=1}^{K}A_{kj}^{2}e^{-\phi_{(j)}(t'-t)}$,

(iii) when $k'=k$, $\text{cov}(\eta_{i}^{k}(t), \eta_{i'}^{k}(t') = (\bLam_{i}^{(k)})^{T}\bLam_{i'}^{(k)}\sum_{j=1}^{K}A_{kj}A_{k'j}e^{-\phi_{(j)}(t'-t)}$.

So, the marginal distribution of $Y_{i}^{(k)}(t)$ is $Y_{i}^{(k)}(t) \sim N(\mu_{i}^{(k)}(t), (\Delta_{i}^{(k)})^{T} \Delta_{i}^{(k)} \sum_{j=1}^{K} A_{kj}^{2})$.  Therefore, with $\mu_{i}^{(k)}(t) = \bX_{i}(t)^{T} \bbet_{i}^{(k)}$, for clustering across $i$, the levels of the $\bX_{i}(t)$ do not matter.  We only ask whether $\bbet_{i}^{(k)} = \bbet_{i'}^{(k)}$?

The marginal distribution for $\bY_{i}(t)$ is also directly available, i.e., $\bY_{i}(t) \sim MVN(\bmu_{i}(t), \bSigma_{ii} + \bD)$ where $\bD$ is a  $K \times K$ diagonal matrix with $(\bD)_{kk} = \tau^{2(k)}$ and $\bSigma_{ii}$ also $K \times K$ with entries $(\bSigma_{ii})_{kk'} $ equal to (ii) above with $t' = t$.

The marginal distribution for the entire $nK \times 1$ vector $\bY(t) = \left(
                                                                          \begin{array}{c}
                                                                            \bY_{1}(t) \\
                                                                            \bY_{2}(t) \\
                                                                            . \\
                                                                            . \\
                                                                            . \\
                                                                            \bY_{n}(t) \\
                                                                          \end{array}
                                                                        \right)$ is

$$\bY(t) \sim MVN(\bmu(t), \bSigma + \bI \otimes \bD)$$
where $\bI$ is the $n \times n$ identity matrix, $\bD$ is as above, and $\bSigma$ is $nK \times nK$ with $K \times K$ blocks such that block $\bSigma_{ii'}$ has entries, $(\bSigma_{ii'})_{kk'}$ equal to (i) above.

Finally, the conditional distribution, $[\bY(t')|\bY(t)]$ can be obtained from the joint distribution of $(\bY(t), \bY(t'))$.  That is, $\left(
                                  \begin{array}{c}
                                    \bY(t) \\
                                    \bY(t') \\
                                  \end{array}
                                \right) \sim  MVN$  $( \left(
                                                    \begin{array}{c}
                                                      \bmu(t) \\
                                                      \bm(t') \\
                                                    \end{array}
                                                  \right)$,$ \left(
                                                             \begin{array}{cc}
                                                               \bSigma + \bI \otimes \bD & \bSigma_{\bY(t),\bY(t')} \\
                                                               \bSigma_{\bY(t), \bY(t')} & \bSigma + \bI \otimes \bD \\
                                                             \end{array}
                                                           \right))$
where  $\bSigma_{\bY(t), \bY(t')}$ is $nK \times nK$ symmetric with $K \times K$ blocks and block $(\bSigma_{\bY(t), \bY(t')})_{ii'}$ has entry $k,k'$ as \eqref{eq:dep}.  Therefore,

$$[\bY(t')|\bY(t)] \sim MVN(\bmu(t') + \bSigma_{\bY(t), \bY(t')}(\bSigma + \bI \otimes \bD)^{-1}(\bY(t) - \bmu(t)), $$

$$(\bSigma + \bI \otimes \bD) - \bSigma_{\bY(t), \bY(t')}(\bSigma + \bI \otimes \bD)^{-1}\bSigma_{\bY(t), \bY(t')}).$$

\section{Gibbs Sampling}\label{app:gibbs}

Here, we provide the Gibbs sampling for a slightly more general model than that presented between (\ref{eq:timevar})-(\ref{eq:ar_ou}). In addition, we use the prior parameters denoted in Section \ref{sec:priors_fitting}. As we do with the Mexico City data, we allow limiting cluster based on a subset of the explanatory variables
\begin{equation*}
\bY^m_{i}(t) =  \bG^m_{i} \bZ_i(t) + \bB^m_{i} \bX_i(t) + \boeta_{i}(t) + \bepsilon_{i}(t),
\end{equation*}
where the $k$th row of $\bG^m_{i}$ is $\bgamma_{im}^{(k)}$. Let $\theta | \cdots$ denote the posterior conditional distribution for $\theta$. With the exception of decay parameters of the exponential covariance functions, $\phi_{(j)}$, the posterior conditional distributions of all model parameters are conveniently available and presented below. Let $c$ indicate cluster and $n_{cm}$ be the number of clusters in time segment $m$. In addition, we constrain the coregionalization matrix $\bA$ to be lower diagonal and index matrix elements $A_{kl}$ for $k < l$.

\begin{align*}
\bgamma_{im}^{(k)} | \cdots &\sim \mathcal{N}\left( \Sigma_{\gamma_{im}^{(k)}} \mu_{\gamma_{im}^{(k)}},\Sigma_{\gamma_{im}^{(k)}} \right) \\
\bbeta_{cm}^{(k)} | \cdots &\sim \mathcal{N}\left( \Sigma_{\beta_{cm}^{(k)}} \mu_{\beta_{cm}^{(k)}},\Sigma_{\beta_{cm}^{(k)}} \right) \\
\Lambda^{(k)}_{i} &\sim \mathcal{N}\left( \Sigma_{\Lambda^{(k)}_{i} } \mu_{\Lambda^{(k)}_{i} },\Sigma_{\Lambda^{(k)}_{i} } \right) \\
A_{kl} &\sim \mathcal{N}\left( v_{A_{kl}}\mu_{A_{kl}},v_{A_{kl}} \right) \\
\tau^2_k &\sim IG\left( a_{\tau^2_k}^*, b_{\tau^2_k}^* \right) \\
\xi_{im} &\sim \text{Multinomial}\left( \pi_{im} \right) \\
\nu^{(k)}(t) &\sim \mathcal{N}\left( \Sigma_{\nu^{(k)}(t)} \mu_{\nu^{(k)}(t)} , \Sigma_{\nu^{(k)}(t)} \right) \\
\end{align*}

where

\begin{align*}
\Sigma_{\gamma_{im}^{(k)}} &= \left( \frac{1}{\tau^2_k} \sum_{t \in m} \bz_{i}(t) \bz_{i}(t)^T + V^{1}_{\gamma_m^{(k)}} \right)^{-1} \\
\mu_{\gamma_{im}^{(k)}} &= \frac{1}{\tau^2_k} \sum_{t \in m} \bz_{i}(t) \left( Y_{i}^{(k)}(t) - \bx_i(t)^T \beta^{(k)}_{im} - \eta_{i}^{(k)}(t) \right) + V^{-1}_{\gamma_m^{(k)}} m_{\gamma_m^{(k)}} \\
\Sigma_{\beta_{cm}^{(k)}} &= \left(\frac{1}{\tau^2_k}\sum_{i \in c} \sum_{t \in m} \bx_i(t) \bx_{i}(t)^T  + V_{\beta^{k}_m}^{-1} \right)^{-1} \\
\mu_{\beta_{cm}^{(k)}} &= \sum_{i \in c} \sum_{t \in m} \bx_i(t) \left( Y_i^{k}(t) - \bz_i(t)^T \bgamma_m^{(k)} - \eta^{(k)}_i(t) \right) + V^{-1}_{\beta_{m}^{(k)}} m_{\beta_{m}^{(k)}} \\
\Sigma_{\Lambda^{(k)}_{i} }  &= \left( \frac{1}{\tau^2_k} \sum^T_{t = 1} \bnu(t)^T \bA_{k} \bA_{k}^T \bnu(t)  + \bV_{\Lambda_{i}^{(k)} }^{-1} \right)^{-1} \\
\mu_{\Lambda^{(k)}_{i} } &= \frac{1}{\tau^2_k} \sum^T_{t = 1} \nu(t)^T \bA_k \left( Y^{(k)}_i(t) -  \bx_i(t)^T \bbeta_{im}^{(k)} - \bz_i(t)^T \bgamma_{im}^{(k)} \right) \\
v_{A_{kl}} &= \left( \frac{1}{s_{A_{kl}}^2} + \frac{1}{\tau^2_k} \sum^T_{t = 1} (\bnu^{(l)}(t))^T (\Lambda^{(k)})^T \Lambda^{(k)} \bnu^{(l)}(t)  \right)^{-1} \\
\mu_{A_{kl}} &= \frac{m_{A_{kl}} }{s_{A_{kl}}^2} + \frac{1}{\tau^2_k} \sum^T_{t = 1} (\bnu^{(l)}(t))^T (\Lambda^{(k)})^T  \left( Y^{(k)}_i(t) -  \bx_i(t)^T \bbeta_{im}^{(k)} - \bz_i(t)^T \bgamma_{im}^{(k)} \right) \\
a_{\tau^2_k}^* &= a_{tau^2_k} + \frac{T \times n_s }{2}  \\
 b_{\tau^2_k}^* &= b_{tau^2_k} + \frac{1}{2} \sum_{t =1}^T \sum_{i = 1}^{n_s} \left( Y_{i}^{(k)}(t) - \bx_i(t)^T \bbeta_{im}^{(k)} - \bz_i(t)^T \bgamma_{im}^{(k)} - \eta_i^{(k)}(t) \right)^2
\end{align*}

To update the Dirichlet process parameters, we use a collapsed sampler \citep{escobar1995}. To do this, we find the marginal distribution of the data for the at location $i$ and time block $m$, integrating over $\bbeta^{k}_{im}$ with respect to its prior distribution. This allows to compute the probability of creating a new group or cluster. Here, we let $\mu^{(k)}_{i,-\beta}(t) = \bz_i(t)^T \bgamma_{im}^{(k)} + \eta_i^{(k)}(t)$ and $n_{c,-i}$ be the number stations in the $c$th if we exclude the $i$th location. Then, the relative probabilities of each cluster is
\begin{equation*}
\pi_{im} \propto \begin{cases}
n_{c,-i} \prod_{t \in m} \prod_{k = 1}^K \mathcal{N}\left( Y^{(k)}_i(t) |\mu^{(k)}_{i,-\beta}(t) + \bx_i(t)^T \beta_{cm}^{(k)}, \tau^2_k \right) \\
\alpha \prod_{k=1}^K \mathcal{N}\left( Y^{(k)}_{im} | \Sigma^{(k)}_{dp} \mu^{(k)}_{dp}, \Sigma_{dp} \right),
\end{cases}
\end{equation*}
where
\begin{align*}
\Sigma^{(k)}_{dp} &= \left[ \frac{1}{\tau^2_k} \bI -  \frac{ X_{im} }{\tau^2_k} \left( \frac{X_{im}^T X_{im}}{\tau^2_k} + V_{\beta}^{-1} \right) \frac{ X_{im}^T }{\tau^2_k} \right]^{-1} \\
\mu^{(k)}_{dp} &= \frac{\mu^{(k)}_{im,-\beta}}{\tau^2_k} - \frac{ X_{im} }{\tau^2_k} \left( \frac{X_{im}^T X_{im}}{\tau^2_k} + V_{\beta}^{-1} \right) \left( \bV^{-1}_{\beta} m_{\beta} + \frac{1}{\tau^2_k} X_{im}^T \mu_{im}^{(k)} \right)  .
\end{align*}
Using the Woodbury matrix formula, we can efficiently computate the inverse of the covariance matrix and the determinant.

For the GP parameters, we define a few terms that allow us to make an $r \times 1$ update. Each functional factor is defined using $K$ independent GPs. First, we define two diagonal matrices $D_{v}^{(k)} = \text{diag}\left( \frac{1}{1 - e^{-2 \phi_1^{(k)}}},...,\frac{1}{1 - e^{-2 \phi_{r}^{(k)}}} \right)$ and $D_{\phi}^{(k)} = \text{diag}\left(  e^{- \phi_1^{(k)}},...,e^{- \phi_r^{(k)}} \right)$. In our example, $D_{v}^{(k)} = D_{v}^{(k')}$ for all $k$ and $k'$. Then, for $  1 < t < T$,
\begin{align*}
\Sigma_{\nu^{(k)}(t)} &= \left( D_v^{(k)} + D_\phi^{(k)} D_v^{(k)} D_\phi^{(k)} + \sum^K_{l = 1} \frac{A_{lk}^2}{\tau_l^2} (\Lambda^{(l)})^T\Lambda^{(l)} \right)^{-1} \\
\mu_{\nu^{(k)}(t)} &= D_v D_\phi \left( \nu^{(k)}(t-1) + \nu^{(k)}(t+1) \right) + \\ &\sum^K_{l = 1} \frac{ A_{lk} }{ \tau^2_l } ( \Lambda^{(l)})^T \left( Y^{(l)}(t) - X(t) \bbeta_{im}^{(l)} - Z(t) \bgamma_{im}^{(l)} - \sum_{k' \neq k} A_{lk'} \Lambda^{(l)} \nu^{(k')}(t)  \right)
\end{align*}
When, $t = 1$, $D_v^{(k)} + D_\phi^{(k)} D_v^{(k)} D_\phi^{(k)}$ becomes $\bI + D_v^{(k)}$ and there are no past terms in the mean. For $t = T$, $\Sigma_{\nu^{(k)}(t)}$ excludes $ D_\phi^{(k)} D_v^{(k)} D_\phi^{(k)}$, as well as future terms in the mean.

If desired, coefficients, $\bgamma_{im}^{(k)}$ and $\bbeta_{cm}^{(k)}$, can be specified hierarchically or dynamically. In this case, the Gibbs sampler would include prior distributions and updates for $m_{\gamma^{(k)}_m} $, $m_{\beta^{(k)}_m} $,$V_{\gamma^{(k)}_m}$, and $V_{\beta^{(k)}_m} $.

\section{Sensitivity Analysis}\label{app:sensitivity}

\begin{figure}
\begin{center}
\includegraphics[width=0.45\textwidth]{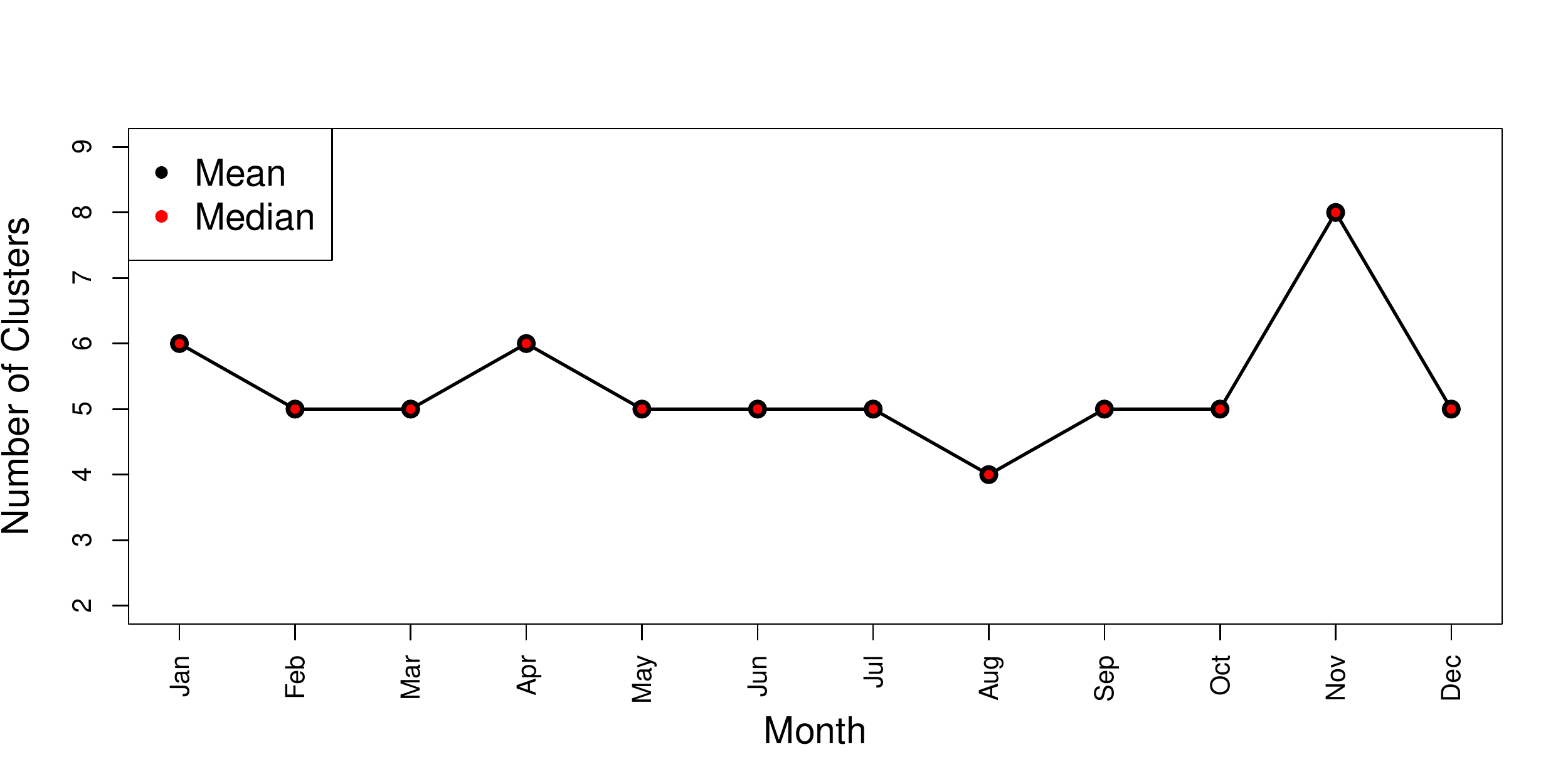}
\includegraphics[width=0.45\textwidth]{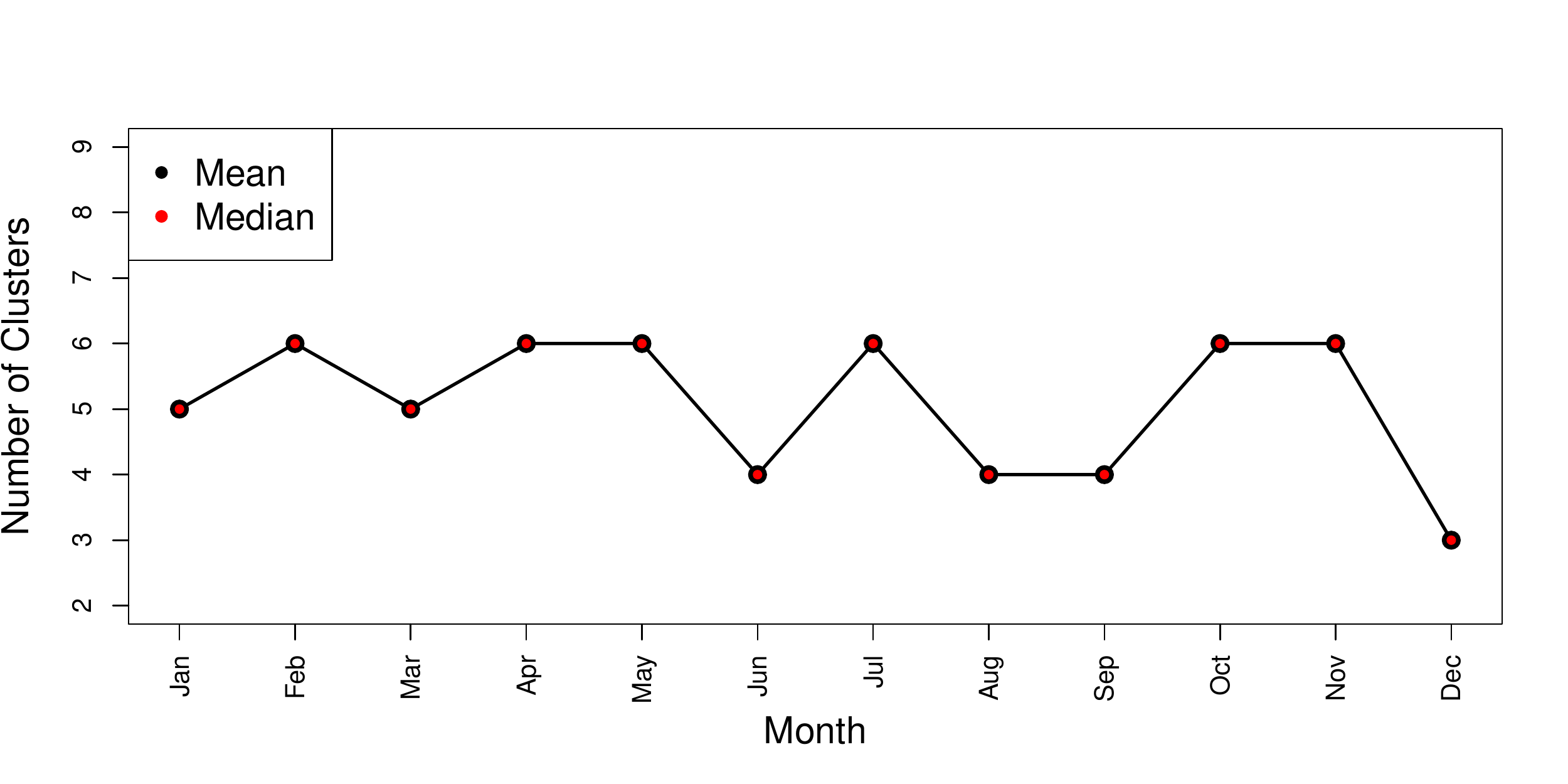}
\includegraphics[width=0.45\textwidth]{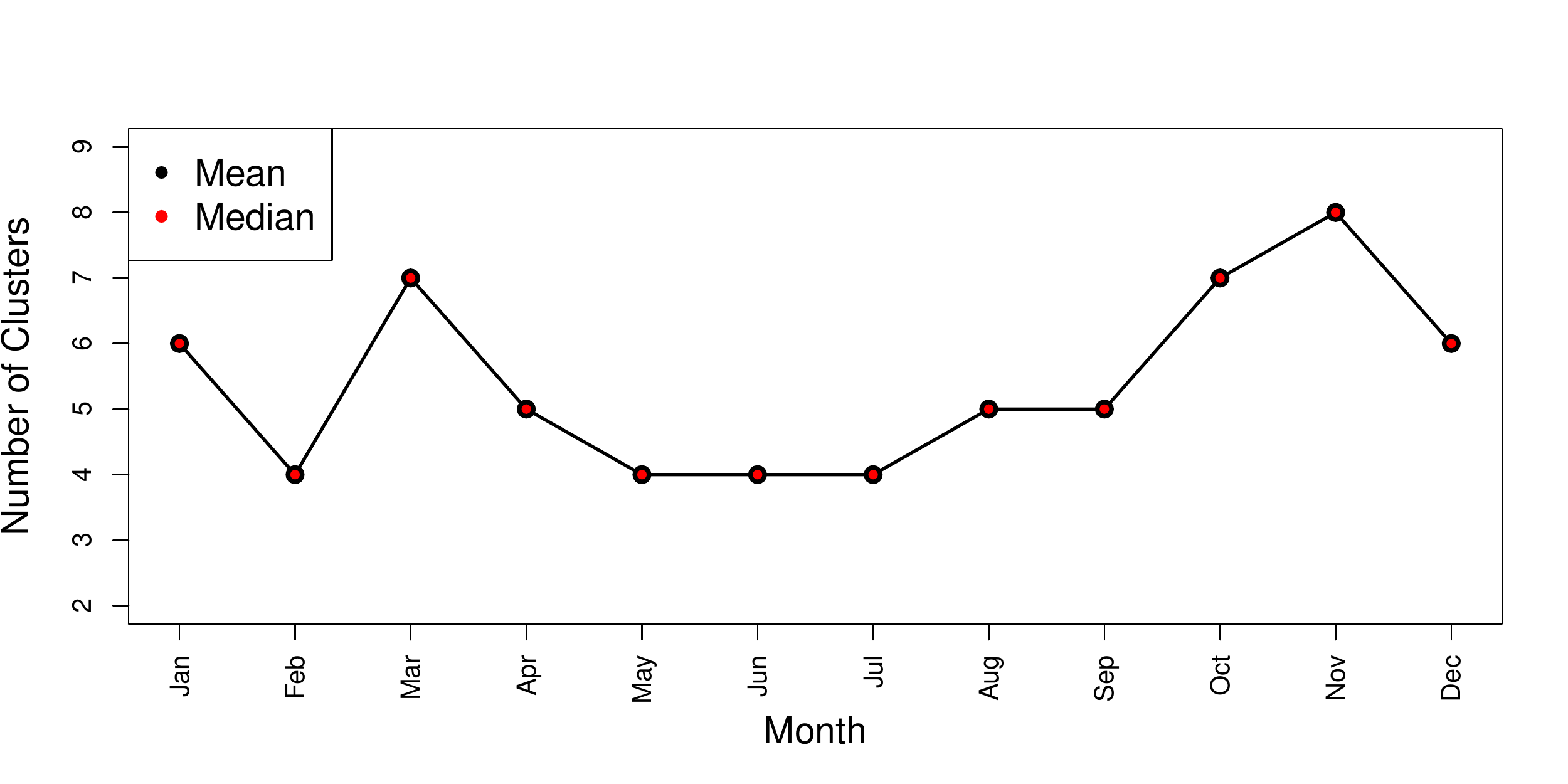}
\includegraphics[width=0.45\textwidth]{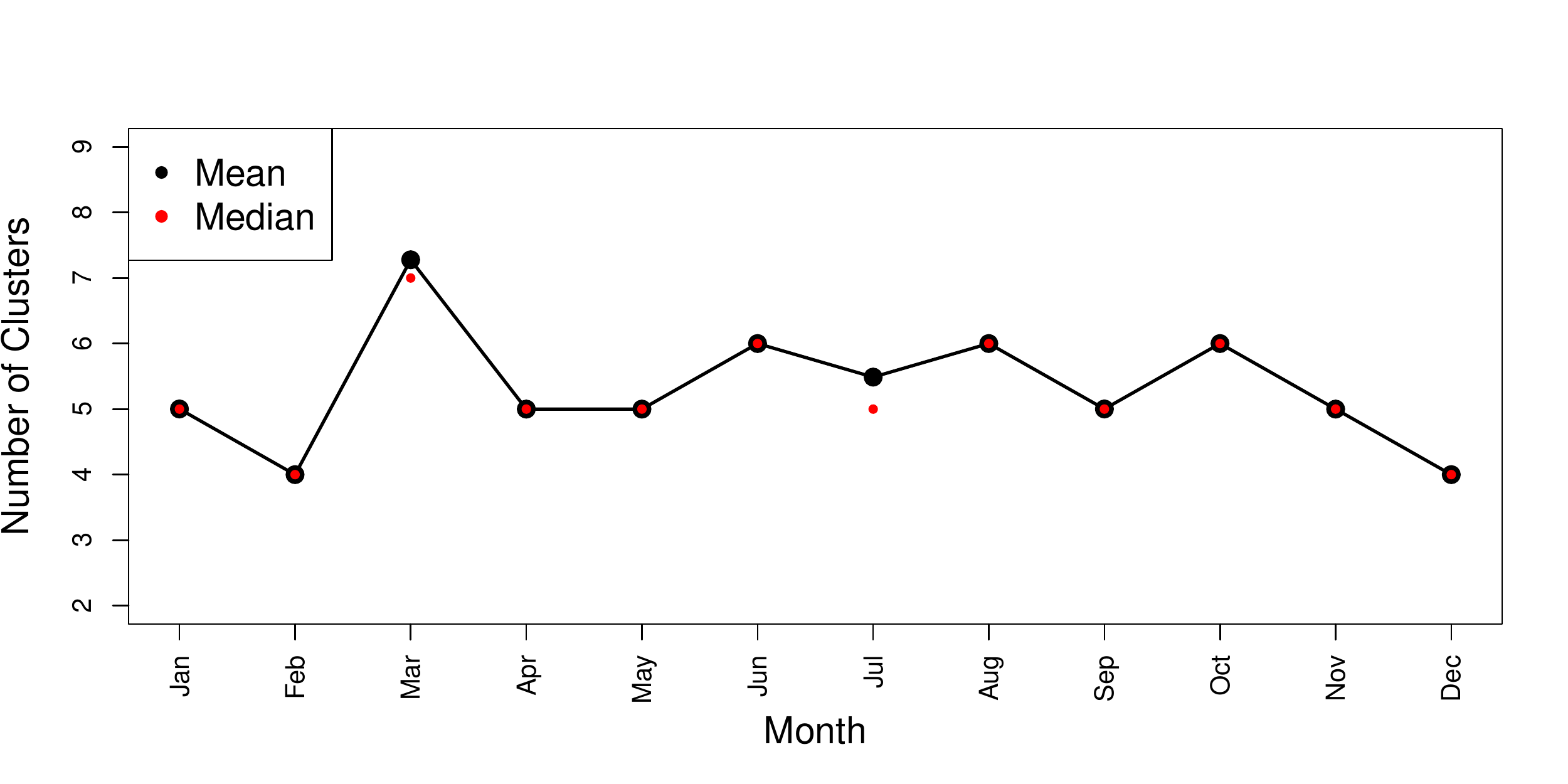}
\includegraphics[width=0.45\textwidth]{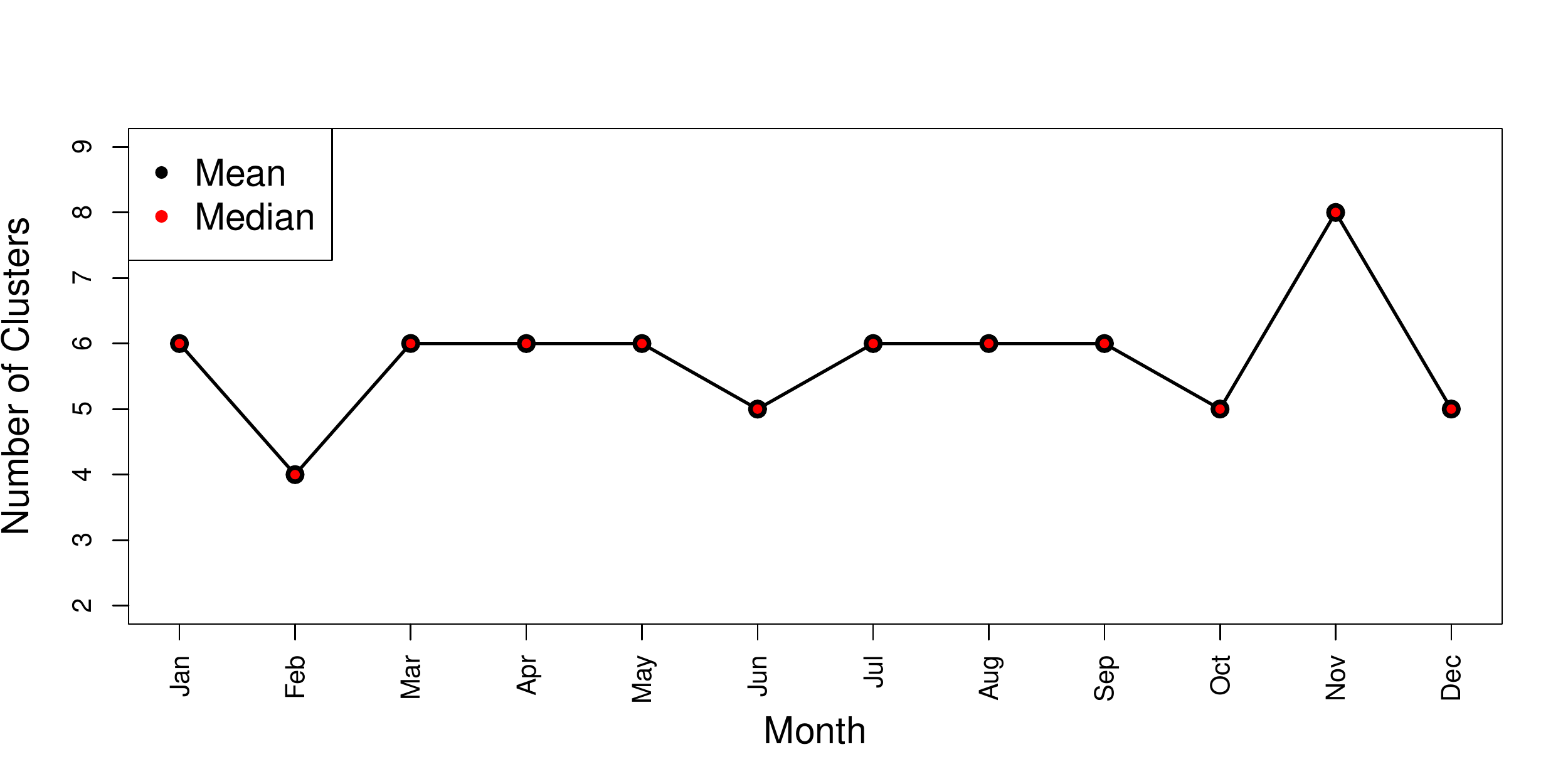}
\includegraphics[width=0.45\textwidth]{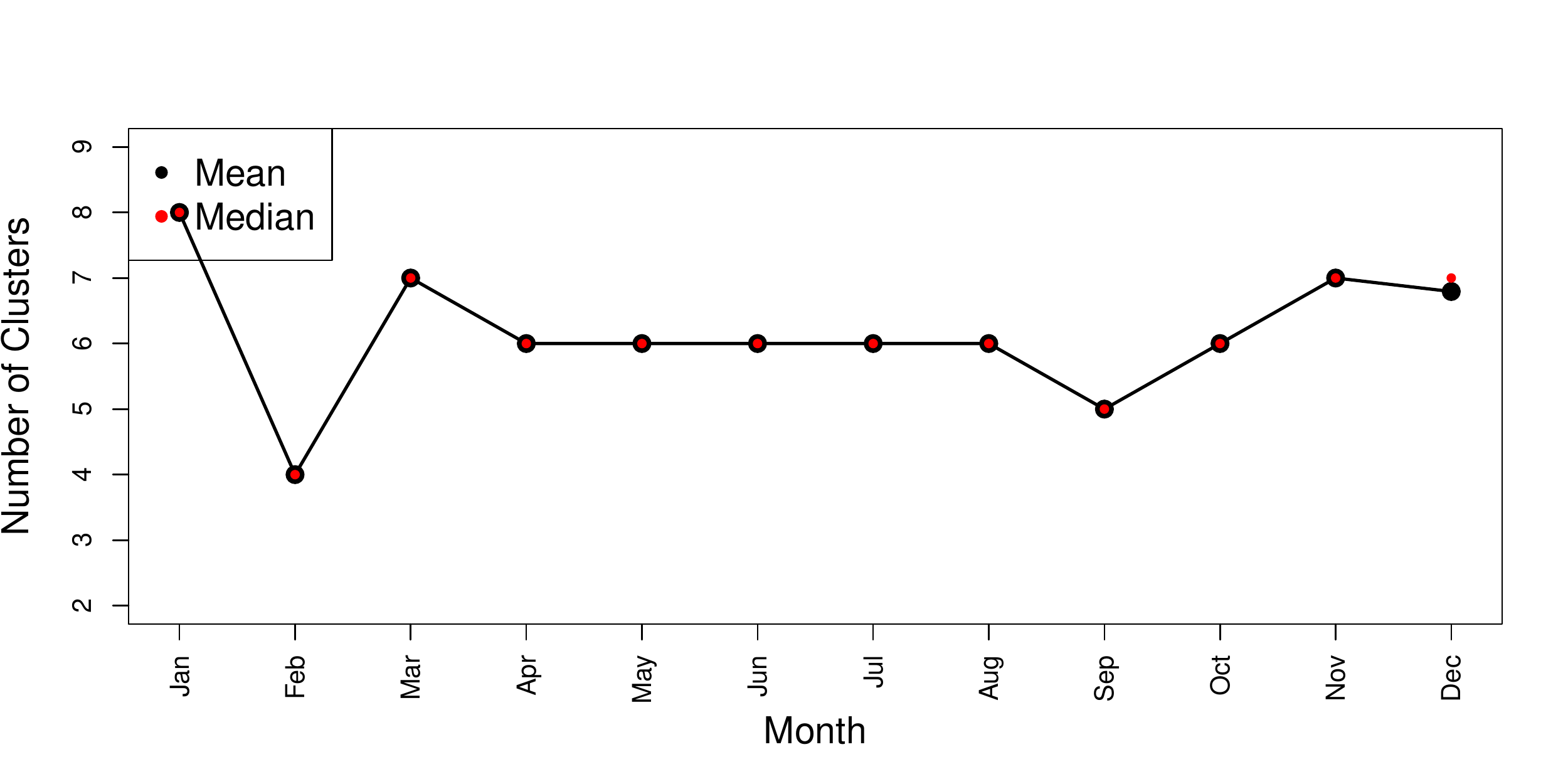}
\includegraphics[width=0.45\textwidth]{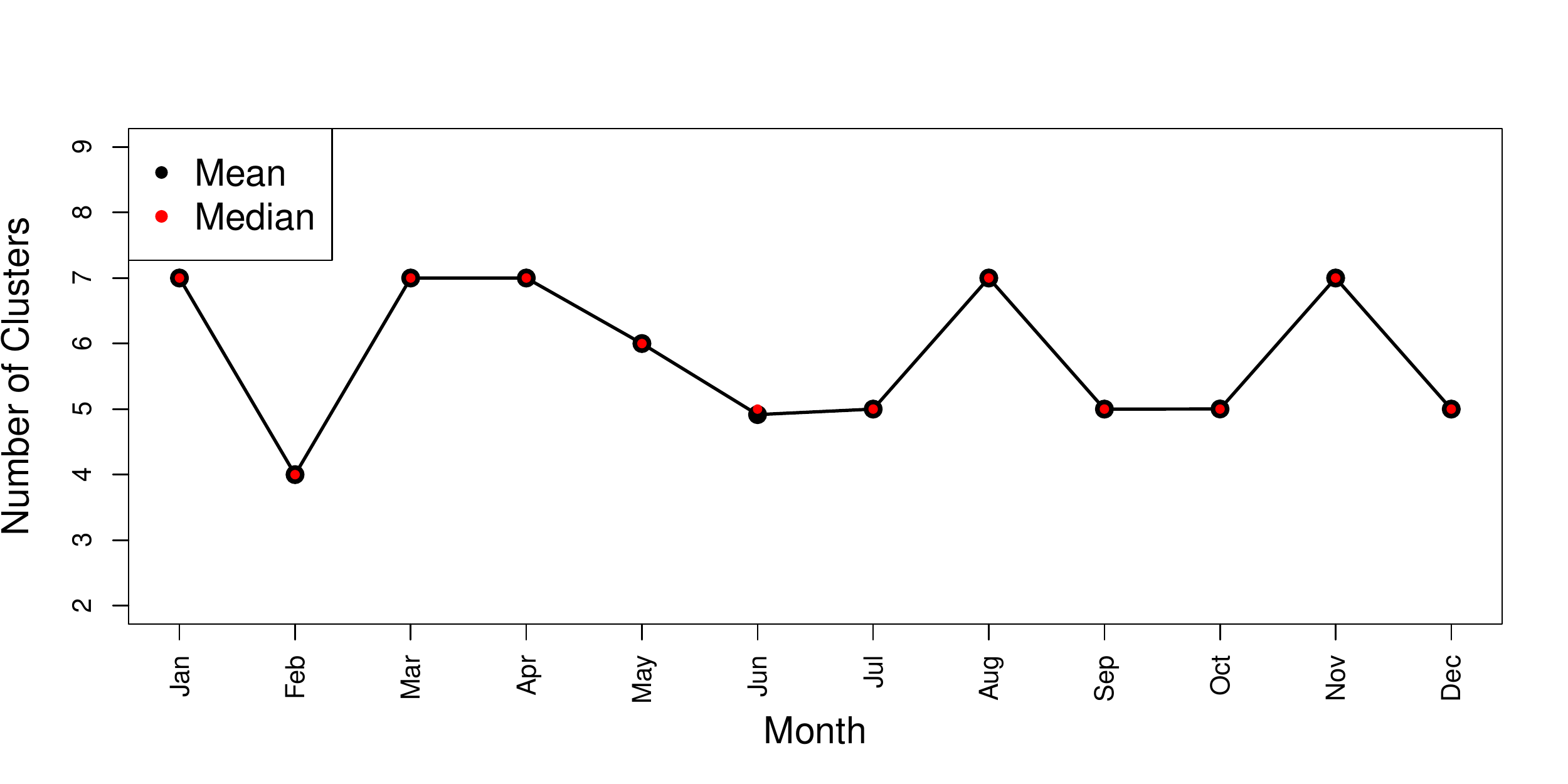}
\includegraphics[width=0.45\textwidth]{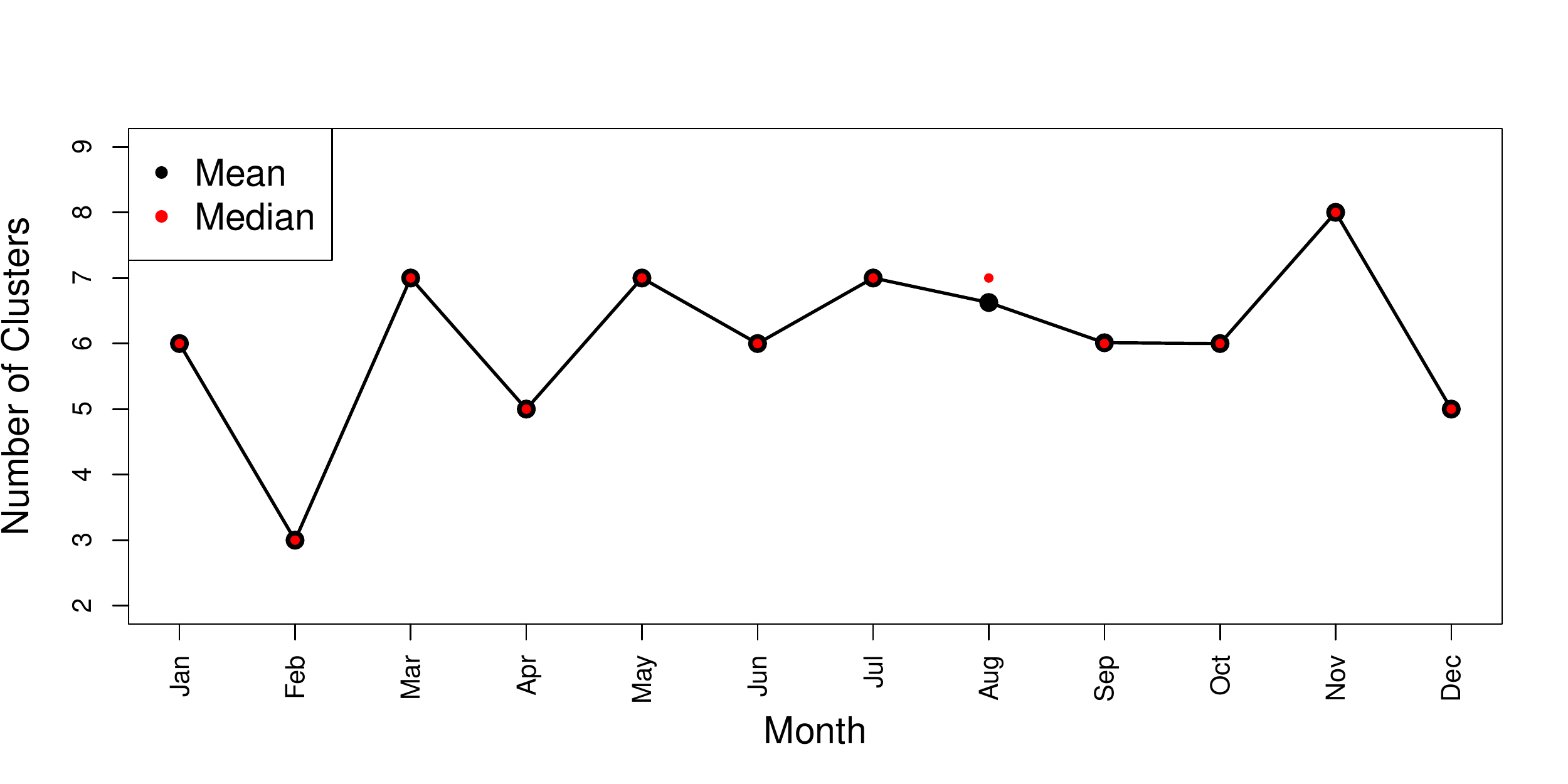}
\includegraphics[width=0.45\textwidth]{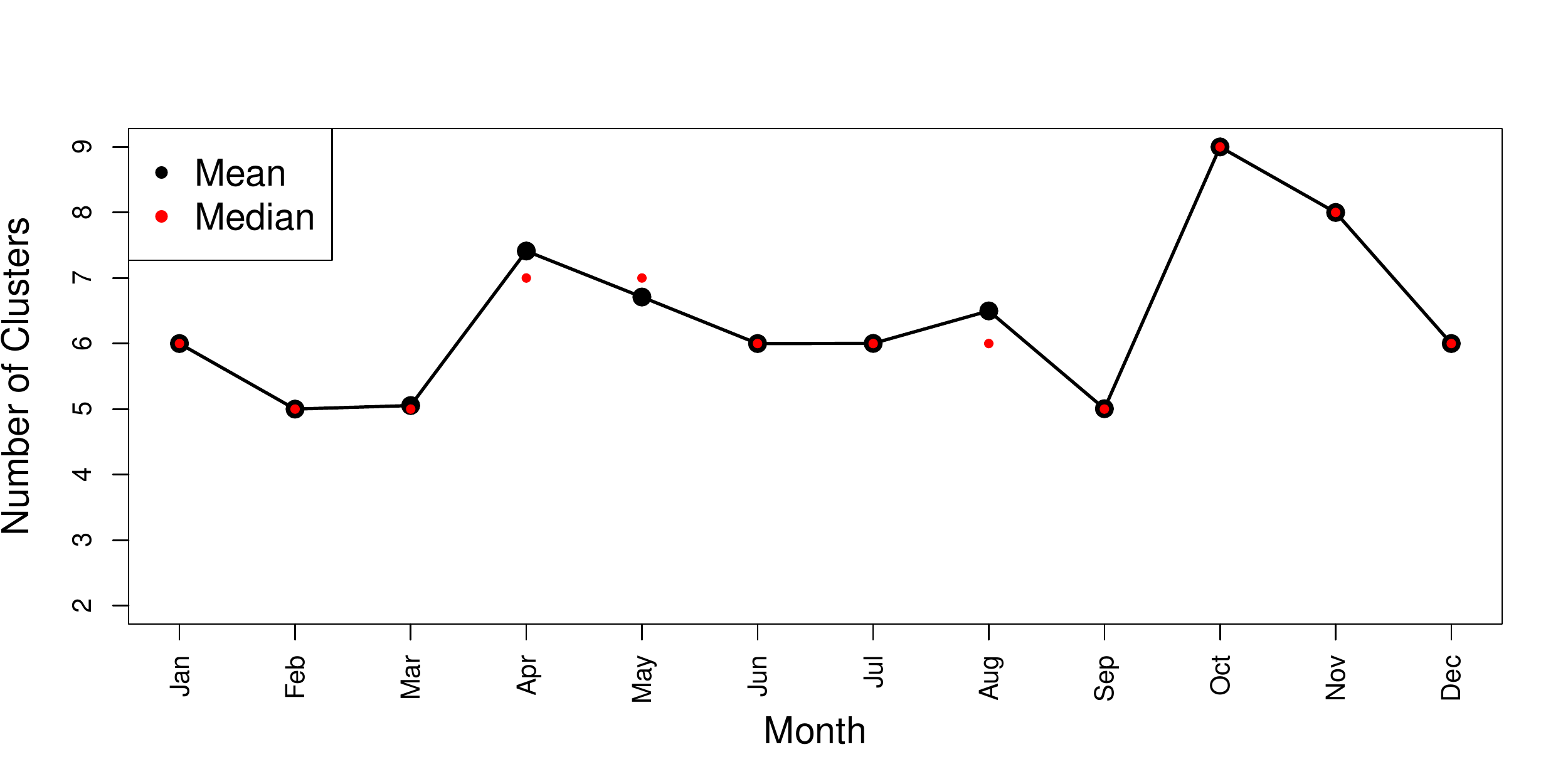}
\includegraphics[width=0.45\textwidth]{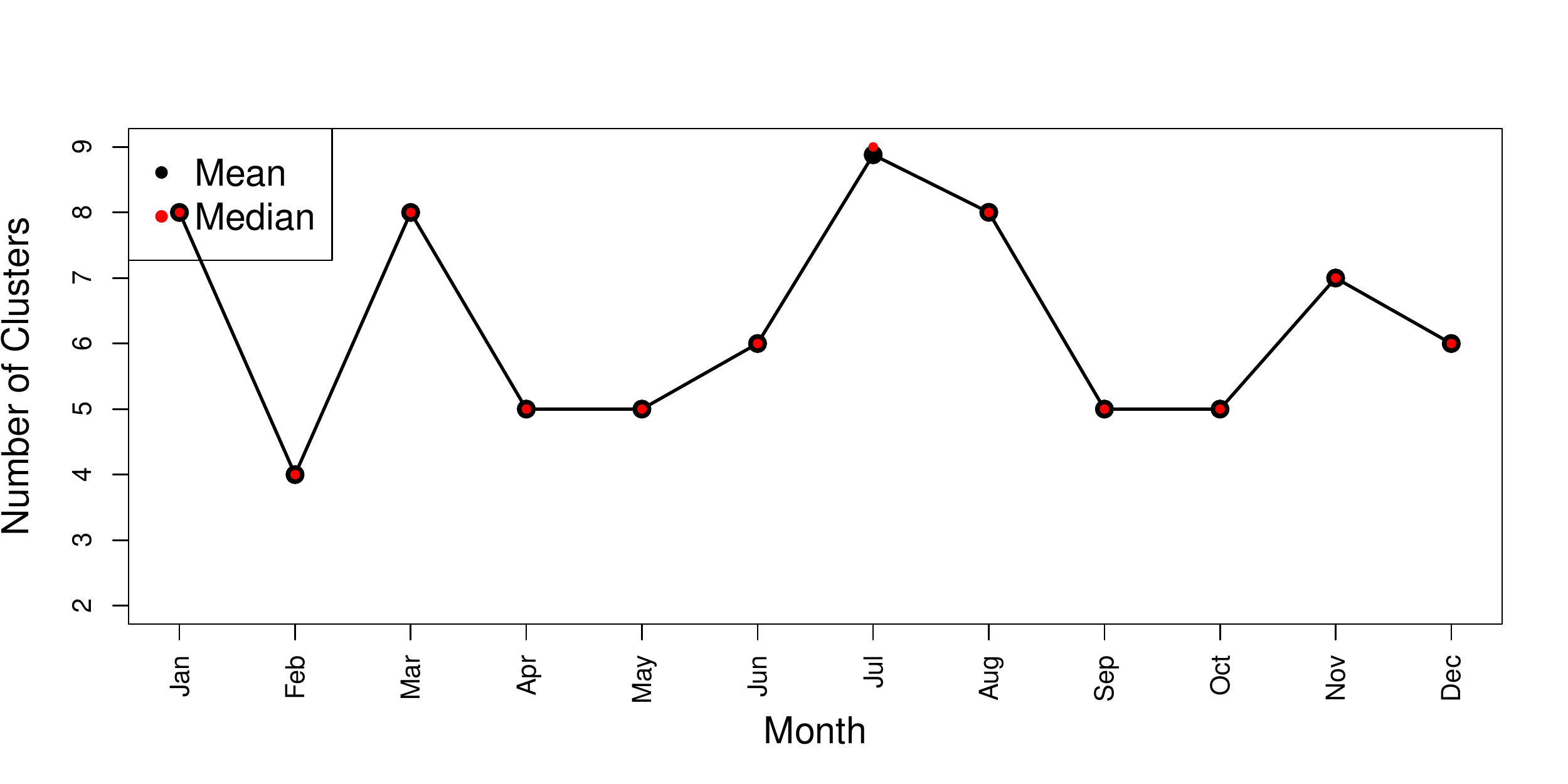}
\end{center}
\caption{Sensitivity in mean and median number of clusters with different concentration parameters $\alpha$ and different prior distributions on the noise variance ${\tau^2}^O$ and ${\tau^2}^{PM}$. In the left columns all noise variance prior distributions are IG($1,1$), while those on the right have IG($101,10$) prior distributions. From top to bottom, we use concentration parameters (1/1000,1/10,1,10,1000). }\label{fig:sensitivity}
\end{figure}

Here, we take up the sensitivity of our model to the concentration parameter $\alpha$ and the prior distributions on the noise parameters $\tau^2_k$ and $\bbeta^{(k)}_{im}$. First, we present how sensitive clustering is to prior specifications of noise parameters $\tau^2$ and concentration parameters. To do this, we supply the mean and median number of clusters as a function of month for many combinations of prior distributions for the error variance and the fixed concentration parameter $\alpha$ (See Figure \ref{fig:sensitivity}). We provide the average number of clusters averaged over month and over iterations of the Gibbs sampler in Table \ref{tab:sensitivity_mean}. In general, with larger concentration parameters, the number of clusters increases. With more diffuse prior distributions on $\tau^2$ parameters, we generally had weaker clustering (more clusters); however, this depended somewhat upon the concentration parameter.

\begin{table}[H]
\centering
\begin{tabular}{l|rrrrr}
  \hline
$\alpha =  $ & $\frac{1}{1000}$ & $\frac{1}{10}$ & $1$ & $10$ & $1000$ \\
  \hline
$\tau^2 \sim IG(1,1)$ & 5.33 & 5.42 & 5.75 & 5.83 & 6.39 \\
 $\tau^2 \sim IG(101,10)$ & 5.08 & 5.31 & 6.15 & 6.05 & 6.32 \\
   \hline
\end{tabular}
\caption{Mean number of clusters, averaged over months and iterations}\label{tab:sensitivity_mean}
\end{table}

\begin{figure}[H]
\begin{center}
\includegraphics[width=0.7\textwidth]{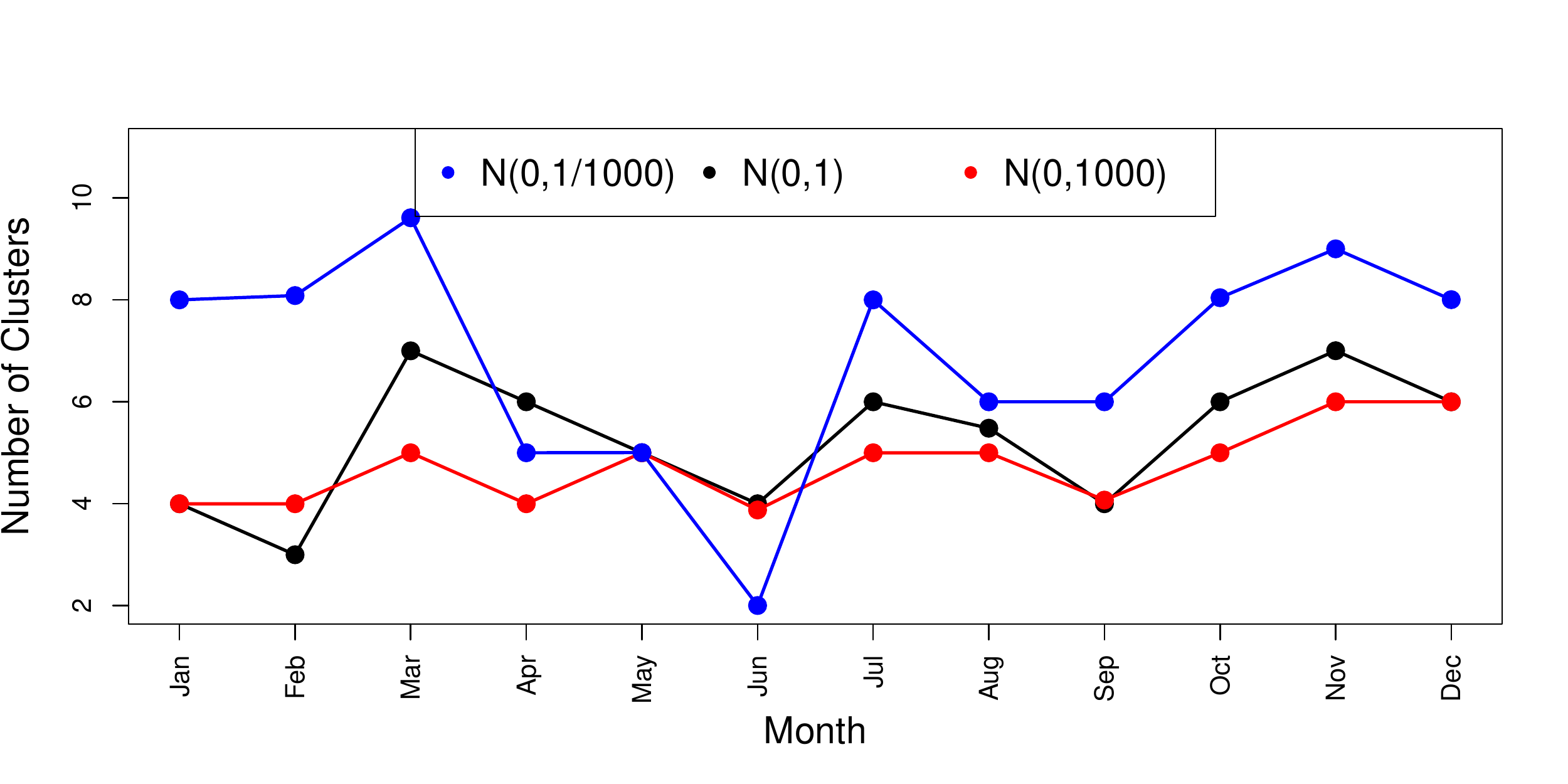}
\end{center}
\caption{Number of clusters over months for different base measures.}\label{fig:beta_sensitivity}
\end{figure}

When creating a new cluster, the marginal likelihood, integrated over the base measure, must be high (See Appendix \ref{app:gibbs}). With a diffuse base measure, the marginal likelihood is lower and fewer clusters are expected. Because the variance of the base measure affects clustering, we explore how different base measures change the observed number of clusters. We plot the average number of clusters over month of the year in Figure \ref{fig:beta_sensitivity}. For the highest variance base measure, $\mathcal{N}\left(\bzero, 10^3 \bI \right)$, we observe 4.75 clusters, averaged over month and iteration.  For base measures of $\mathcal{N}\left(\bzero,  \bI \right)$ and $\mathcal{N}\left(\bzero, 10^{-3} \bI \right)$, we observe the average number of clusters to be 5.29 and 6.90, respectively.  As expected, when the variance of the base measure is small, we have more clusters, on average. On the other hand, a very diffuse base measure leads to fewer clusters. This results jibes with our intuition. Given that we have centered and scaled the outcomes and covariates, a very diffuse base measure for the regression coefficients seems unreasonable.

\begin{figure}[H]
\begin{center}
\includegraphics[width=0.7\textwidth]{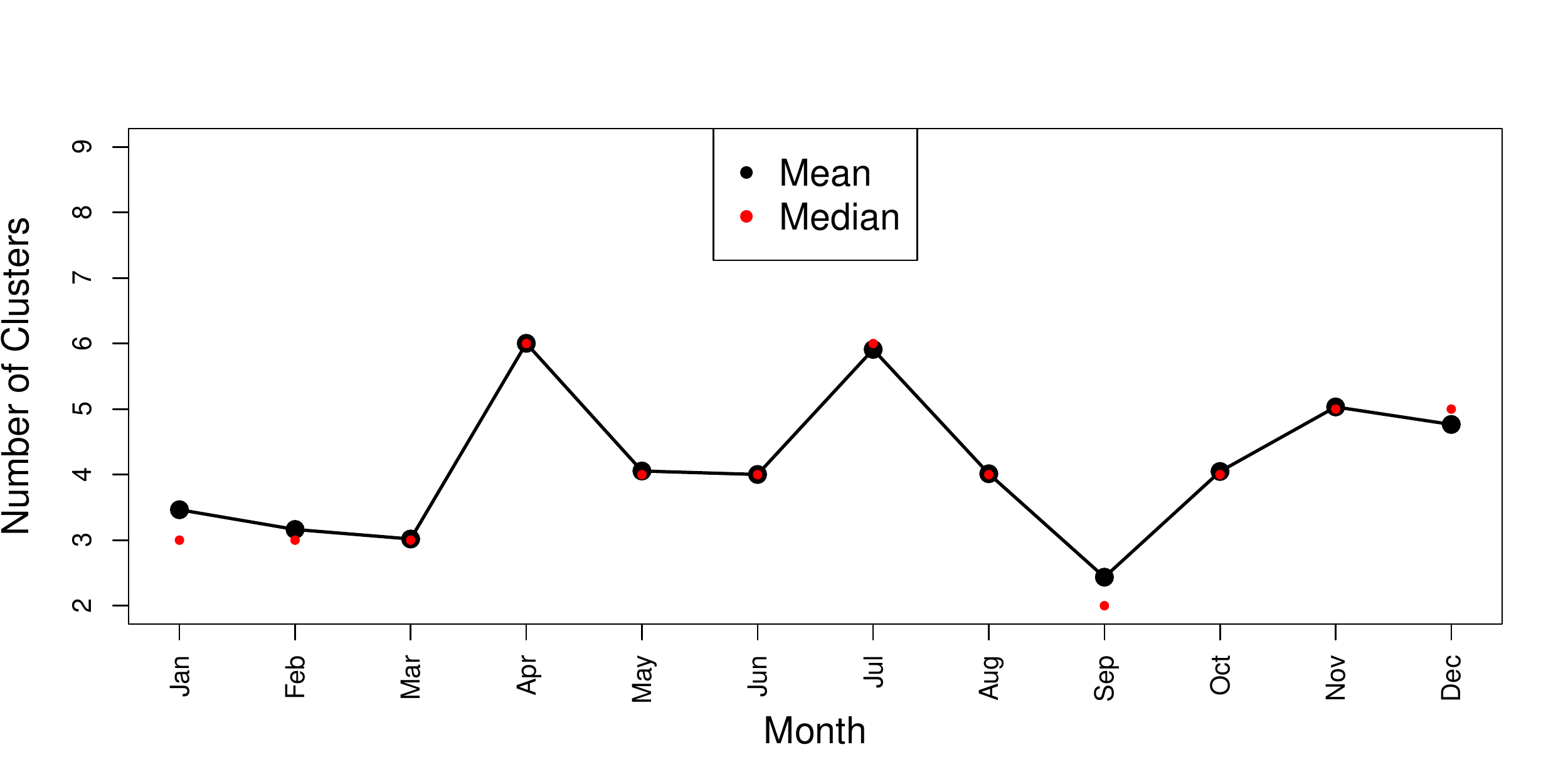}
\end{center}
\caption{Sensitivity in mean and median number of clusters for ozone's regression coefficients over months.}\label{fig:ind_sensitivity}
\end{figure}

We also briefly discuss our decision to jointly cluster regression coefficients for ozone and \pmt. 
In preliminary modeling, we noticed that regression coefficients for \pmt would not cluster with reasonable values of $\alpha$. Ozone, on the other hand, would cluster. When the concentration parameter for both ozone and \pmt was 1, the mean number of clusters averaging over month and iterations of the Gibbs sampler were 3.98 and 24 for ozone and \pmt, respectively. Therefore, we argue that the joint clustering we have discussed in Section \ref{sec:results} is driven by clustering in ozone coefficients. In Figure \ref{fig:ind_sensitivity}, we plot the mean and median number of ozone clusters as a function of month.

\bibliographystyle{apalike}
\bibliography{refs_mexico}

\end{document}